\documentclass{article}

\pdfoutput=1
\usepackage{graphicx,amsmath,amssymb,epsf} \usepackage{latexsym,bm,
  slashed} \usepackage{xcolor} \definecolor{dark}{rgb}{0.10,0.2,0.3}
\definecolor{magenta}{rgb}{0.7,0.1,0.3}
\definecolor{purpure}{rgb}{0.5,0.15,0.3}
\usepackage[font=small,format=plain,labelfont=bf,up,textfont=it,up]{caption}
\usepackage{hyperref, cite} \hypersetup{colorlinks, citecolor=blue,
  filecolor=blue, linkcolor=magenta,
  urlcolor=purpure,hyperfootnotes=true,pdftex} 

\usepackage{subcaption}
\usepackage{siunitx}
\usepackage{physics}

\newcommand{\pM}[1][]{p_{\perp\text{min}#1}}
\newcommand{\pT}[1][]{p_{\perp#1}}
\newcommand{\pobs}[1][]{\langle \pT\rangle_{\text{mini}#1}}
\newcommand{\yobs}[1][]{\langle \Delta y \rangle_{\text{mini}#1}}
\newcommand{\nM}[1][]{N_{\text{mini}#1}}
\newcommand{\Y}[0]{\Delta Y}
\newcommand{\Ym}[0]{\Delta Y_\text{min}}

\title{Multijet event shape variables for Mueller--Navelet~jet~topologies}

\author{C. Baldenegro$^{\, 0}$, G. Chachamis$^{\, 2,3}$, M. Kampshoff$^{\, 1,4}$, M. Klasen$^{\, 4}$, \\ G. J. Milhano$^{\, 2,5}$,
C. Royon$^{\, 1}$, A. Sabio Vera$^{\, 6,7}$\\ 
\\
\small $^0$ Dipartimento di Fisica, Sapienza Universit{\' a} di Roma, Piazzale Aldo Moro, 2, 00185 Rome, Italy\\
\small $^1$ Department of Physics \& Astronomy, The University of Kansas, KS 66047, USA.\\
\small $^2$ Laborat{\' o}rio de Instrumenta\c{c}{\~ a}o e F{\' \i}sica Experimental de Part{\' \i}culas (LIP),\\
\small Av. Prof. Gama Pinto, 2, P-1649-003 Lisboa, Portugal.\\
\small $^3$ Departamento de Estad\'istica, Inform\'atica y Matem\'aticas, \\
\small Universidad P\'ublica de Navarra-UPNA, 31006 Pamplona, Spain. \\
\small $^4$Institut f{\"u}r Theoretische Physik, Universit{\" a}t M{\"u}nster,\\ 
\small Wilhelm-Klemm-Stra{\ss}e 9, D-48149 M{\"u}nster, Germany.\\
\small $^5$ Departmento de F\'isica, Instituto Superior T\'ecnico (IST),\\
\small Universidade de Lisboa, Av. Rovisco Pais 1, P-1049-001 Lisboa, Portugal.\\
\small $^6$ Instituto de F{\'\i}sica Te{\'o}rica UAM/CSIC, Nicol{\'a}s Cabrera 15, E-28049 Madrid, Spain.\\
\small $^7$ Theoretical Physics Department, Universidad Aut{\' o}noma de Madrid, E-28049 Madrid, Spain.\\} 

\begin{document} 

\maketitle 

\begin{abstract}

This paper presents a new set of multijet event shape variables introduced to further understand the Mueller--Navelet jet topology. This topology consists of having at least one pair of jets with a very large rapidity separation between them, treating additional jet activity inclusively. This multijet topology is expected to shed light on the radiation pattern that is expected in the high-energy limit of the strong interaction. The paper relies on a Monte Carlo event generator analysis. One set of predictions uses the \texttt{BFKLex} event generator, which is based on Balitsky--Fadin--Kuraev--Lipatov (BFKL) perturbative quantum chromodynamics (pQCD) evolution with a resummation of large logarithms of energy at leading-logarithmic accuracy. The \texttt{BFKLex} predictions are compared with a fixed-order next-to-leading order pQCD calculation using \texttt{POWHEG} matched to the parton shower of \texttt{PYTHIA}8 (NLO+PS), which is the standard for NLO generator predictions at the LHC. We find that both approaches lead to compatible results at current LHC energies, assuming the current experimental constraints for the reconstruction of low transverse momentum jets in ATLAS and CMS. This shows the reliability of the BFKL approach at describing the behavior of the strong interaction in the preasymptotic limit of high center-of-mass energies. Differences between the NLO+PS and the BFKL-based approaches are found when the jet multiplicity is increased or when the minimum transverse momentum of the jets is decreased.

\end{abstract}

\vspace*{-150mm}
\hfill MS-TP-24-NN
\vspace*{143mm}

\section{Introduction}

\begin{figure} [!h]
    \centering
    \includegraphics[width=0.5\textwidth]{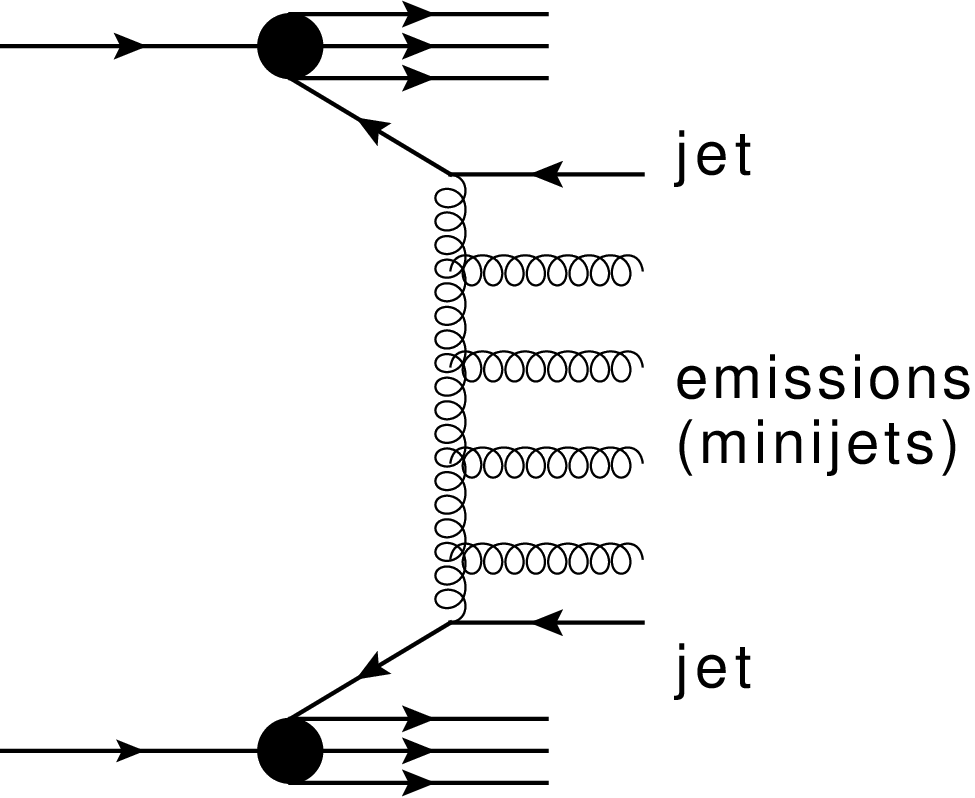}
    \caption{Schematic diagram of multijet production in the high-energy limit of QCD (Mueller--Navelet jet topology). The gluon-ladder diagrams are expected to dominate in this limit.}
    \label{fig:MNdiagram}
\end{figure}

One of the goals of the Large Hadron Collider (LHC) at CERN is to understand the emergent properties of the strong force, some of which are expected from quantum chromodynamics (QCD), the theory of the strong interaction. In the limit where the center-of-mass energy of the hard scattering is much larger than other hard scales, known as the high-energy limit, the contribution from multigluon emissions compensate the smallness of the strong coupling $\alpha_S$ via large logarithms of energy. In practice, those can be resummed with the Balitsky--Fadin--Kuraev--Lipatov (BFKL) approach~\cite{Lipatov:1985uk,Balitsky:1978ic,Kuraev:1977fs,Kuraev:1976ge,Lipatov:1976zz,Fadin:1975cb,Marquet:2007xx,Navelet:1996jx,Marquet:2004tw,Peschanski:2004vw,Kepka:2006cg,Kepka:2006xe,Marquet:2012ra,Kepka:2010hu,Marquet:2005vp}. A similar resummation is needed for the small-$x$ evolution of parton distribution functions (PDFs). To test the emergent properties expected from the strong force in the high-energy limit, numerous processes have been explored. One such process is the production of jets in a Mueller--Navelet (MN) topology~\cite{Mueller:1986ey}, which is a type of dijet event where two high-energy jets are produced with a large separation in rapidity, treating the rest of the jet activity in the event inclusively, as shown schematically in Fig.~\ref{fig:MNdiagram}. One may, for example, analyze the angular correlation between the forward and backward jets, which have been calculated at next-to-leading-logarithmic (NLL) accuracy in the BFKL formalism. In the last two decades, there were numerous studies in the literature of this process, see for example Refs.~\cite{Bartels:2001ge,Bartels:2002yj,Colferai:2010wu,Caporale:2011cc,Angioni:2011wj,Caporale:2012ih,Ducloue:2013hia,Ducloue:2013bva,Caporale:2013uva,Caporale:2014gpa,Celiberto:2015yba,Caporale:2015uva,Mueller:2015ael,Celiberto:2016ygs,Celiberto:2022gji,CMS:2016qng,CMS:2021maw,ATLAS:2011juz}. The LHC measurements that focus on the angular correlation between the forward and backward jets are consistent with predictions calculated using BFKL resummation at NLL accuracy. However, they are also consistent with other calculations that do not have a dedicated resummation of large logarithms of energy \textit{{\`a} la} BFKL. Thus, there is a need to explore other physical observables, perhaps by focusing on the additional jet activity between the forward-backward MN jets in terms of event shape variables.\\

In the collinear factorization scheme, accurate knowledge of the partonic cross section is very important for computing differential distributions. When we shift from using fixed-order matrix elements to employing the BFKL framework for determining the hard scattering part of the collision, it offers an opportunity to compare the two distinct dynamics and learn about their similarities and differences. This comparison depends on other essential components, mainly the PDFs not having significant influence on the specific differential distributions under examination. Even in cases where the observable under study fails to clearly differentiate between the two dynamics, it remains a crucial task as it usually allows us to identify the effect responsible for their similarity and subsequently search for new observables that exhibit greater sensitivity to the underlying dynamics of the hard scattering process. Nevertheless, it could also indicate that the two dynamics yield similar estimates for a given observable, in which case we learn more about their similarities, a work that is much harder to do by analytic means.

In this paper, we present a discussion and a proposal for an improvement of several observables that were introduced by two of the authors in Refs.~\cite{Chachamis:2015ico,deLeon:2021ecb} for the description of multijet final states in events with a MN jet configuration at the LHC. When the rapidity separation of these two tagged jets is very large (each jet close to the acceptance of the detectors), the phase space available for intrajet radiation is enlarged. In turn, this larger phase space lays ground for further opportunities of a distinction between a regime describable in terms of BFKL evolution. We analyze in detail the additional jet activity at central rapidities, i.e., those jets located between the most forward and backward jets used to tag the event as a MN dijet event topology. \\

Jets are clusters of particles that produced by the parton branching and hadronization of a high-virtuality parton produced in the hard scattering. Hereafter, we use the term ``jets'' or ``MN jets'' for the two outermost (in rapidity) jets and  ``minijets'' to refer to the jets located between such MN jets (minijets was also the term used in Ref.~\cite{Mueller:1986ey}). Such minijets can be related to the gluon-initiated radiation from the gluon ladder diagrams in the BFKL approach. One of the main differences between MN jets and minijets is the $\pT$ of the partons that initiate them, as well as their span in rapidity. For a typical MN jet configuration, the MN jets may have a $\pT$ above a given threshold, for instance above 20 GeV, whereas minijets may have a lower $\pT$, typically with values in the range of 1--5 GeV. To take into consideration experimental limitations on the calibration of jets by CMS and ATLAS, as well as resilience to MPI contributions, we consider cases where all jets have at least $\pT > 20$ GeV.\\

In the following, we present a detailed analysis of these observables for events where the most forward and the most backward jets are tagged, while  $\nM$ minijets are present with rapidities $y_i, i=1, \dots N$, fulfilling $y_i > y_{i+1}$. The minimum number of final state jets for the observables studied here is 2 MN jets and $\nM = 3$, whereas the maximum one is 2 MNjets and $\nM=5$. The two-dimensional transverse momenta of the minijets, $k_i$, are not ordered in any specific way. In Sections 2 and 3 we review the BFKL resummation, on which the code {\tt BFKLex} is based, and the fixed-order QCD approach, that were used in this study. In Section 4 we present our numerical results and discuss their implications. Finally, we summarize our findings and suggest some directions for future work.

\section{BFKL and {\tt BFKLex} generator}

Identifying a set of processes, observables, and kinematical regimes where BFKL dynamics can be unambiguously observed has proven to be a challenging task. This is because phenomenological calculations based on matrix elements computed at fixed order along with Dokshitzer-Gribov-Lipatov-Altarelli-Parisi (DGLAP) evolution~\cite{Gribov:1972ri,Gribov:1972rt,Lipatov:1974qm,Altarelli:1977zs,Dokshitzer:1977sg} tend to describe the bulk of the data adequately.

The key idea in the BFKL formalism is that, when the center-of-mass energy $\sqrt{s} \to \infty$, diagrams that contribute with terms of the form {$\alpha_s^n \log^n{\left(s\right)} \sim \alpha_s^n \left(y_A-y_B\right)^n$} yield the dominant numerical contributions to the computation of the cross section. Here, $y_{A}$ and $y_{B}$ are the rapidities of some properly chosen jets (or particles) in the final state, such that their rapidity difference $\Y = y_A - y_B$ is the largest among the particles or jets in the final state. The terms  {$\alpha_s^n \log^n{\left(s\right)} $ can be of order unity, and therefore these diagrams must be resummed in order to stabilize the perturbative expansion and hopefully describe measurements better. In this limit, a decoupling between transverse and longitudinal degrees of freedom takes place, which allows for the evaluation of cross sections in the factorized form
\begin{eqnarray}
\sigma^{\rm LL} &=& \sum_{n=0}^\infty {\cal C}_n^{\rm LL}  \alpha_s^n 
\int_{y_B}^{y_A} d y_1 \int_{y_B}^{y_1} d y_2 \dots \int_{y_B}^{y_{n-1}} d y_n \nonumber\\ 
&=& \sum_{n=0}^\infty \frac{{\cal C}_n^{\rm LL}}{n!} 
\underbrace{\alpha_s^n \left(y_A-y_B\right)^n }_{\rm LL}\,.
\end{eqnarray}
Here, LL stands for the leading-log approximation and the $y_i$ correspond to the rapidities of the emitted particles. The LL BFKL formalism allows one to calculate the coefficients ${\cal C}_n^{\rm LL}$~\cite{Lipatov:1985uk,Balitsky:1978ic,Kuraev:1977fs,Kuraev:1976ge,Lipatov:1976zz,Fadin:1975cb}. The next-to-leading-log approximation (NLL)~\cite{Fadin:1998py,Ciafaloni:1998gs} is much more complicated, since it is sensitive to the running of the strong coupling and to the choice of energy scale in the logarithms. One can parametrize the freedom in the choice of these two scales, respectively, by introducing the constants ${\cal A}$ and ${\cal B}$ in the previous expression,
\begin{eqnarray}
\sigma^{LL+NLL} &=& 
\sum_{n=1}^\infty \frac{{\cal C}_n^{\rm LL} }{n!}  \left(\alpha_s- {\cal A} \alpha_s^2\right)^n \left(y_A-y_B - {\cal B}\right)^n \nonumber\\
&&\hspace{-2.4cm}= \sigma^{\rm LL}  - \sum_{n=1}^\infty   \frac{\left({\cal B}  \, {\cal C}_n^{\rm LL} +  (n-1) \, {\cal A} 
\, {\cal C}_{n-1}^{\rm LL} \right)}{(n-1)!}  \underbrace{ \alpha_s^n 
\left(y_A-y_B\right)^{n-1}}_{\rm NLL} + \dots \,.
\end{eqnarray}
We see that, at NLL, a power in $\log{s}$ is lost w.r.t.\ the power of the coupling. Within this formalism, we can then calculate cross sections using the factorization formula (with $\Y\simeq \ln{s}$)
 \begin{eqnarray}
\sigma (Q_1,Q_2,\Y) = \int d^2 \vec{k}_A d^2 \vec{k}_B \, \underbrace{\phi_A(Q_1,\vec{k}_A) \, 
\phi_B(Q_2,\vec{k}_B)}_{\rm PROCESS-DEPENDENT} \, \underbrace{f (\vec{k}_A,\vec{k}_B,\Y)}_{\rm UNIVERSAL}, 
\end{eqnarray}
where $\phi_{A,B}$ are process-dependent impact factors which are functions of some external scales, $Q_{1,2}$, and some internal momenta for reggeized gluons, $\vec{k}_{A,B}$.  The Green function $f$ is universal. It depends on $\vec{k}_{A,B}$ and on the colliding energy of the process $\sim e^{\Y/2}$ and corresponds to the solution of the BFKL equation. In momentum space, the BFKL equation  at LL reads
\begin{equation}
\omega \, f_\omega\left(\vec{k}_A, \vec{k}_B\right)=\delta^2\left(\vec{k}_A-\vec{k}_B\right)+\int \mathrm{d}^2 \vec{k}\,\,\, \mathcal{K}\left(\vec{k}_A, \vec{k}\right)\,\, f_\omega\left(\vec{k}, \vec{k}_B\right)\,,
\end{equation}
where $\mathcal{K}\left(\vec{k}_a, \vec{k}\right)$ is the BFKL kernel
\begin{equation}
\mathcal{K}\left(\vec{k}_a, \vec{k}\right)=\underbrace{2 \omega\left(-\vec{q}^2\right) \delta^2\left(\vec{k}_a-\vec{k}\right)}_{\mathcal{K}_{\text {virt }}}+\underbrace{\frac{N_c \alpha_s}{\pi^2} \frac{1}{\left(\vec{k}_a-\vec{k}_b\right)^2}}_{\mathcal{K}_{\text {real }}} .
\end{equation}
The solution of the BFKL equation at LL in transverse momentum representation can be written
in an iterative form~\cite{Schmidt:1996fg} as
\begin{eqnarray}
f &=& e^{\omega \left(\vec{k}_A\right) \Y}  \Bigg\{\delta^{(2)} \left(\vec{k}_A-\vec{k}_B\right) + \sum_{n=1}^\infty \prod_{i=1}^n \frac{\alpha_s N_c}{\pi}  \int d^2 \vec{k}_i  
\frac{\theta\left(k_i^2-\lambda^2\right)}{\pi k_i^2} \nonumber\\
&&\hspace{-1.2cm}\int_0^{y_{i-1}} \hspace{-.3cm}d y_i e^{\left(\omega \left(\vec{k}_A+\sum_{l=1}^i \vec{k}_l\right) -\omega \left(\vec{k}_A+\sum_{l=1}^{i-1} \vec{k}_l\right)\right) y_i} \delta^{(2)} 
\left(\vec{k}_A+ \sum_{l=1}^n \vec{k}_l - \vec{k}_B\right)\Bigg\}, 
 \end{eqnarray}
where the gluon Regge trajectory reads
\begin{eqnarray}
\omega \left(\vec{q}\right) &=& - \frac{\alpha_s N_c}{\pi} \log{\frac{q^2}{\lambda^2}}.
\end{eqnarray}
$\lambda$ is a regulator of infrared divergencies. This solution has been studied at length in a series of papers and it served as the basis in order to construct the Monte Carlo event code {\tt BFKLex} which has had multiple 
applications in collider phenomenology and more formal studies~\cite{Chachamis:2022jis,deLeon:2020myv,Chachamis:2015ico,Chachamis:2015zzp,Caporale:2013bva,Chachamis:2012qw,Chachamis:2012fk,Chachamis:2011nz,Chachamis:2011rw}. 
{\tt BFKLex} generates the partonic cross section by combining the gluon Green's function with jet vertices and then convolving it with the PDFs. The resulting event consists of a list of final-state gluons, each accompanied by its four-momentum. The event data can then be fed into a jet clustering algorithm for further analysis.

\section{Fixed-order QCD and parton showers ({\tt POWHEG} + {\tt PYTHIA8}) }

In this section we present our methodology for computing  the NLO QCD predictions for MN jets.
We employ the POWHEG framework~\cite{POWHEGref1,POWHEGref2,POWHEGRef3} to calculate the NLO QCD dijet cross section. We interface POWHEG with \texttt{Pythia}8.3~\cite{PythiaRef} to add parton showers, hadronization and multiple parton interactions (MPIs) in our simulated events. The parameter settings for the event generation are described in Tab.~\ref{tab:setup}, where we give the all the relevant parameters for our analysis (cf.\ also Ref.~\cite{MTPaper}).  \\

\begin{table} 
    \centering
    \caption{Setup for NLO QCD Monte Carlo generators.}
    \label{tab:setup}
    \begin{tabular}{|c|c|c|}
	\hline
	\multicolumn{3}{|l|}{POWHEG parameters}\\
	\hline
	bornktmin&5 GeV&$\pM$ in Born process\\
    bornsuppfact&100 GeV&Suppression factor for high-$\pT$ tail\\
    PDF&NNPDF31\_nlo\_as\_0118&\\
    withnegweights&1&Use negative weights\\
    doublefsr&1&Emitter harder than emitted\\
	\hline
	\multicolumn{3}{|l|}{\texttt{Pythia} 8 parameters}\\
	\hline 
	ISR&on&Initial-state radiation\\
    FSR&on&Final-state radiation\\
    MPI&off&Multi-parton interaction\\
    Tune& pp:14 &Monash 2013, base for CP5\\
    POWHEG:pThard&2&Veto threshold, recommended\\
    POWHEG:pTemt&0&Hardness, recommended \cite{Matching}\\
	\hline
    \end{tabular}
\end{table}

In order to avoid  soft singularities that can arise within the POWHEG framework, we set
the infrared cutoff to \SI{5}{\giga\electronvolt} (\texttt{bornktmin 5}). Since POWHEG produces unweighted events with statistics that follows the $\pT$ distribution of the cross section, low-$\pT$ events were suppressed by a factor of $\pT^2/[\pT^2+(\SI{100}{\giga\electronvolt})^2]$ in order to obtain enough event statistics at higher $\pT$ (\texttt{bornsuppfact 100}). We also use the NNPDF31\_nlo\_as\_0118 PDF sets \cite{NNPDF:2017mvq} by default. Note that the gluon and quark densities do not vary much between different PDFs in the kinematical domain that we use. Furthermore, we note that negative-weighted events were used (\texttt{withnegweights 1}), as recommended for dijet production, as well as the POWHEG routine to reduce the impact of single large-weight events (\texttt{doublefsr 1}). \\

Regarding the settings for \texttt{Pythia}8, we leave initial- and final-state radiation enabled by default in the standard way for inclusive jet production. The impact of the MPIs on our results was minimal. Therefore, we generated our events with MPIs switched off. We used the CP5 \texttt{Pythia}8 tune based on the Monash 2013 tune \cite{Skands:2014pea} as is used by the CMS Collaboration at LHC energies\cite{CMS:2019csb}. \\

Since the NLO matrix elements for the hard scattering part are computed with \texttt{POWHEG} and subsequently interfaced with \texttt{Pythia}8, we need to use matching to avoid over- or undercounting some phase-space regions. The ambiguity in the \texttt{POWHEG}-\texttt{Pythia} matching arises because the scales corresponding to real emissions are slightly different in the two generators: \texttt{POWHEG} passes a value (\texttt{SCALUP}) that can be compared to \texttt{Pythia}8's $\pT$, but the two definitions aren't exactly the same. Hence, different procedures exist to reconcile these differences. 

The matching procedure we followed is known as vetoed showers. In this approach, a separation scale is computed to replace the \texttt{POWHEG} scale and subsequently is compared to the assumed hardness of the \texttt{Pythia}8 emissions. The shower starts at the beam energy, but emissions for which the hardness is larger than the separation scale are suppressed. The separation scale is chosen here as the minimum $\pT$ out of every possible final state particle pair. This choice produces lower multiplicities and preserves more correlations between jets and partons (\texttt{POWHEG:pThard 2}), while the hardness is calculated as the $\pT$ of the emitted parton with respect to the emitting one (\texttt{POWHEG:pTemt 0}).

\section{Results}

\begin{figure} [!h]
    \centering
    \includegraphics[width=0.80\textwidth]{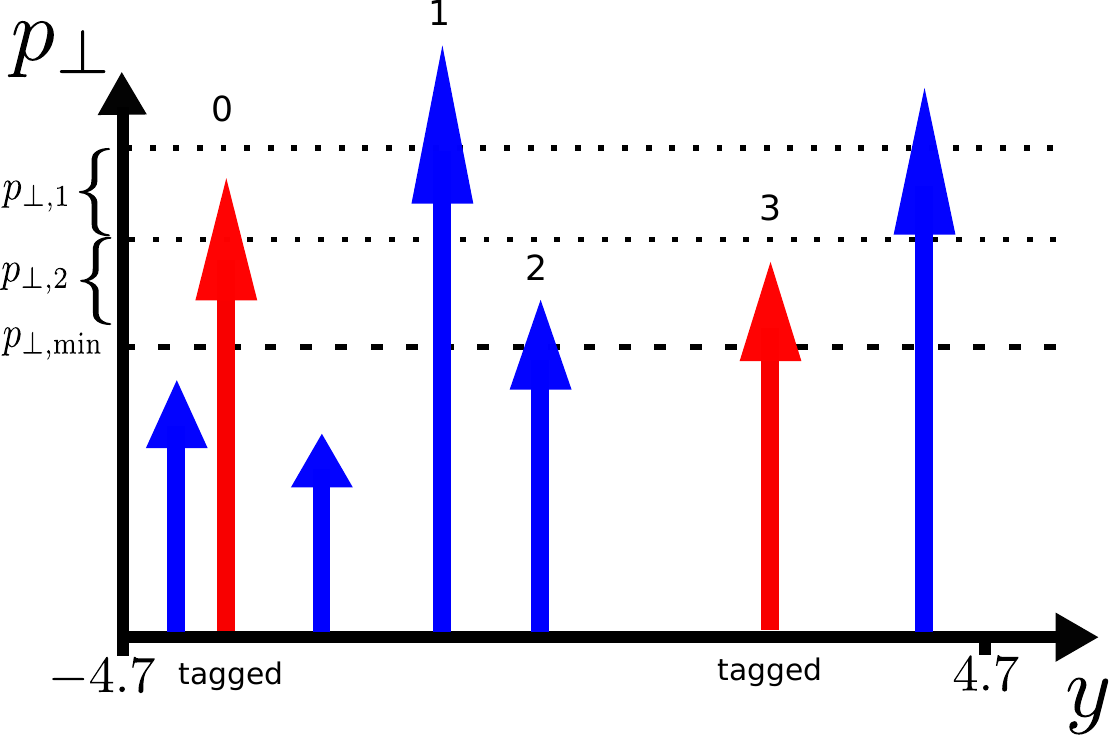}
    \caption{Example of a MN event. The MN jets (red) are 0 and 3, since this pair of jets fulfills the $\pT$ cuts ($\pT[,0]\in[\SI{30}{GeV},\SI{40}{GeV}]\;,\pT[,3]\in[\SI{20}{GeV},\SI{30}{GeV}]$) and has the largest rapidity separation of all possible pairs. The leftmost jet has a $\pT$ value too small to be tagged and is discarded. The third jet does not make the $\pM$ cut and is discarded as well. The rightmost jet is not considered, since it lies outside the two tagged jets. The only jets passing the $\pM$ cut between the tagged jets 0 and 3 are numbers 1 and 2, so this event is classified as a 2-minijet event, $\nM=2$.}
    \label{fig:cutex}
\end{figure}

In this section, we discuss the comparison between the NLO QCD and BFKL predictions for the different observables depending on the minijets produced between the two MN jets. We use the anti-$k_T$ jet algorithm \cite{Cacciari:2008gp} with $R=0.5$ to reconstruct the jets and minijets. As an example, for practical comparisons between NLO QCD and BFKL calculations we use asymmetric requirements on the MN jet transverse momenta, $\pT[,1]\in[\SI{30}{GeV},\SI{40}{GeV}],\;\pT[,2]\in[\SI{20}{GeV},\SI{30}{GeV}]$\cite{AsymCuts}, and rapidities similar to the typical detector acceptances for the ATLAS and CMS experiments, $y_i\in[-4.7,4.7]$. Any jet radiation in rapidities between the MN jets with a $\pT>\SI{20}{\giga\electronvolt}$ (or $\pM>\SI{10}{\giga\electronvolt}$ for a lower $\pT$ selection study) is considered a minijet. An example of a MN event with two minijets is shown in Fig~\ref{fig:cutex}. \\

To compute the minijet observables defined in the next subsections, the rapidities of the jets and minijets in an event are shifted so that the backward MN jet has $y=0$. The minijets are then sorted with ascending rapidity, $y_i<y_{i+1}$. The MN jets themselves are not considered for the computation of the minijets observables~\cite{Chachamis:2015ico}. Furthermore, all the differential distributions shown in the following subsections have been self-normalized unless stated otherwise. \\

For the observables that we discuss in this section, it is expected that differences between the BFKL and NLO QCD predictions will increase with $\nM$. This is because the NLO QCD calculation using {\tt POWHEG} is limited to $2\rightarrow 3$ processes, with  additional radiation generated from {\tt Pythia}. Thus, the higher the number of minijets, the higher the percentage coming from collinear emissions, and the less likely a BFKL-like topology becomes. 

\subsection{Dijet production cross sections}

\begin{figure}
\begin{center}
\includegraphics[width=0.49\textwidth]{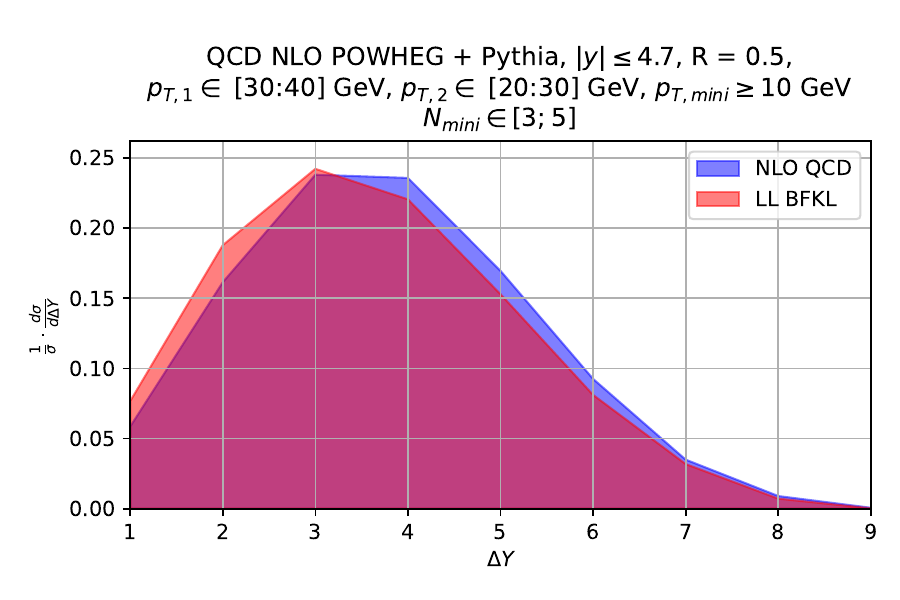}
\includegraphics[width=0.49\textwidth]{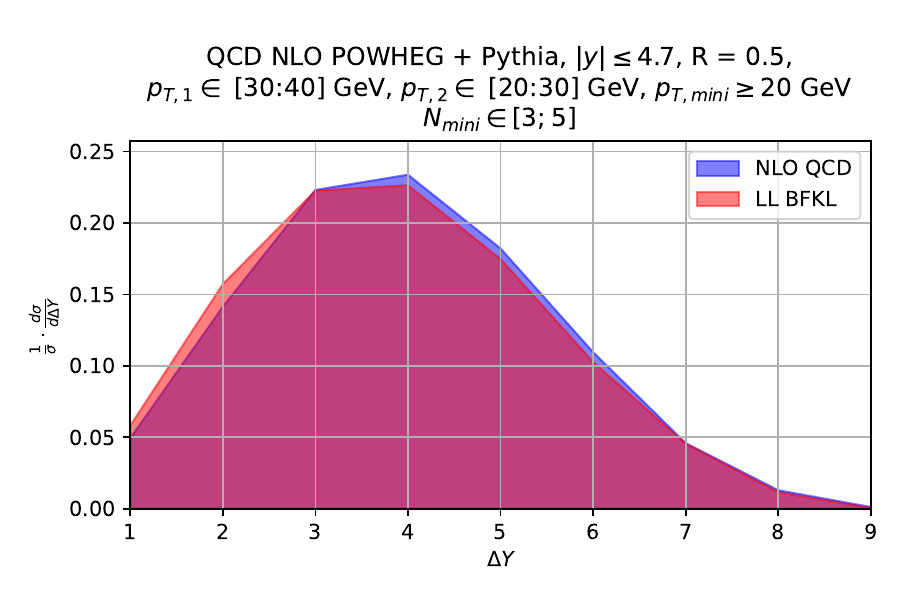}
\caption{Normalized inclusive MN jets distributions as a function of rapidity difference between the MN jets $\Delta Y$ for $\pM=\SI{10}{GeV}$ (left) and $\pM=\SI{20}{GeV}$ (right) for NLO+PS simulation (blue) and BFKL (red). The events that contribute to these distributions have final state jet multiplicity $ \nM \left(= 3, 4, 5\right) + 2 \,(\text{MN jets})$, that is, 5, 6 and 7}.

\label{dSigmadYfull}
\end{center}
\end{figure}

The predicted cross section for inclusive dijet production for NLO QCD in the chosen $\pT$ and $y$ kinematical domain is \SI{1.7}{\micro\barn}. The first experimental observable is the inclusive dijet cross section as a function of rapidity separation of the MN jets $\Y$ and the minijet multiplicity $\nM$ for the event. Fig.~\ref{dSigmadYfull} shows the cross section $d\sigma/d\Y$ for two values of $\pM$. Interestingly, both approaches predict a slightly larger value of $\Y$ for larger $\pM$. Regarding the fixed-order NLO QCD prediction, one can interpret this by assuming that a larger rapidity separation between the MN jets opens up the needed phase space for a third hard minijet to be produced from the matrix elements part. The number of minijets likely to pass the $\pT$ requirement is thus higher at larger $\Y$. For both values of $\pM$, the BFKL prediction peaks at slightly lower values of $\Y$, but the difference is more pronounced for the lower value. \\

\begin{figure}
\begin{center}
\includegraphics[width=0.49\textwidth]{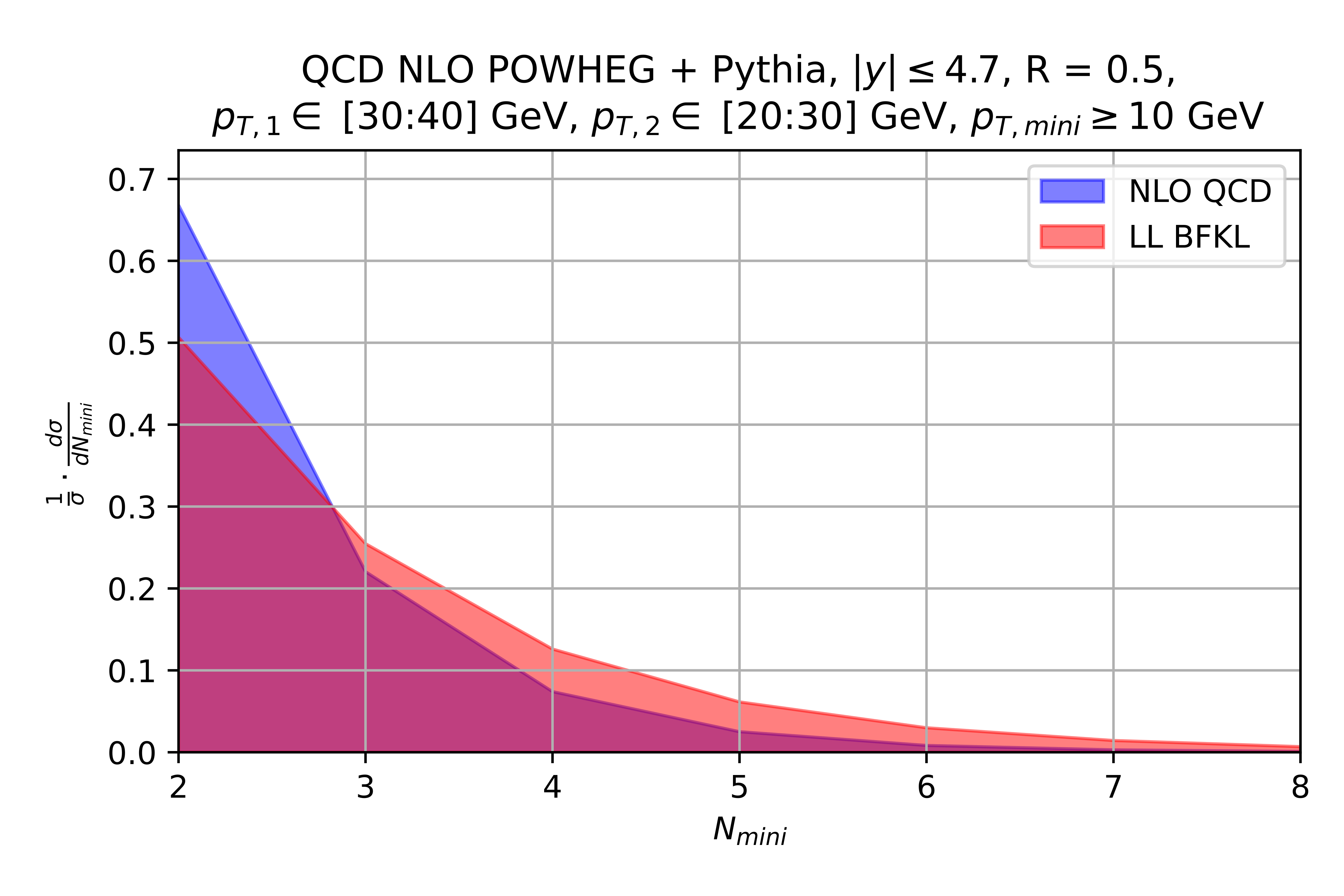}
\includegraphics[width=0.49\textwidth]{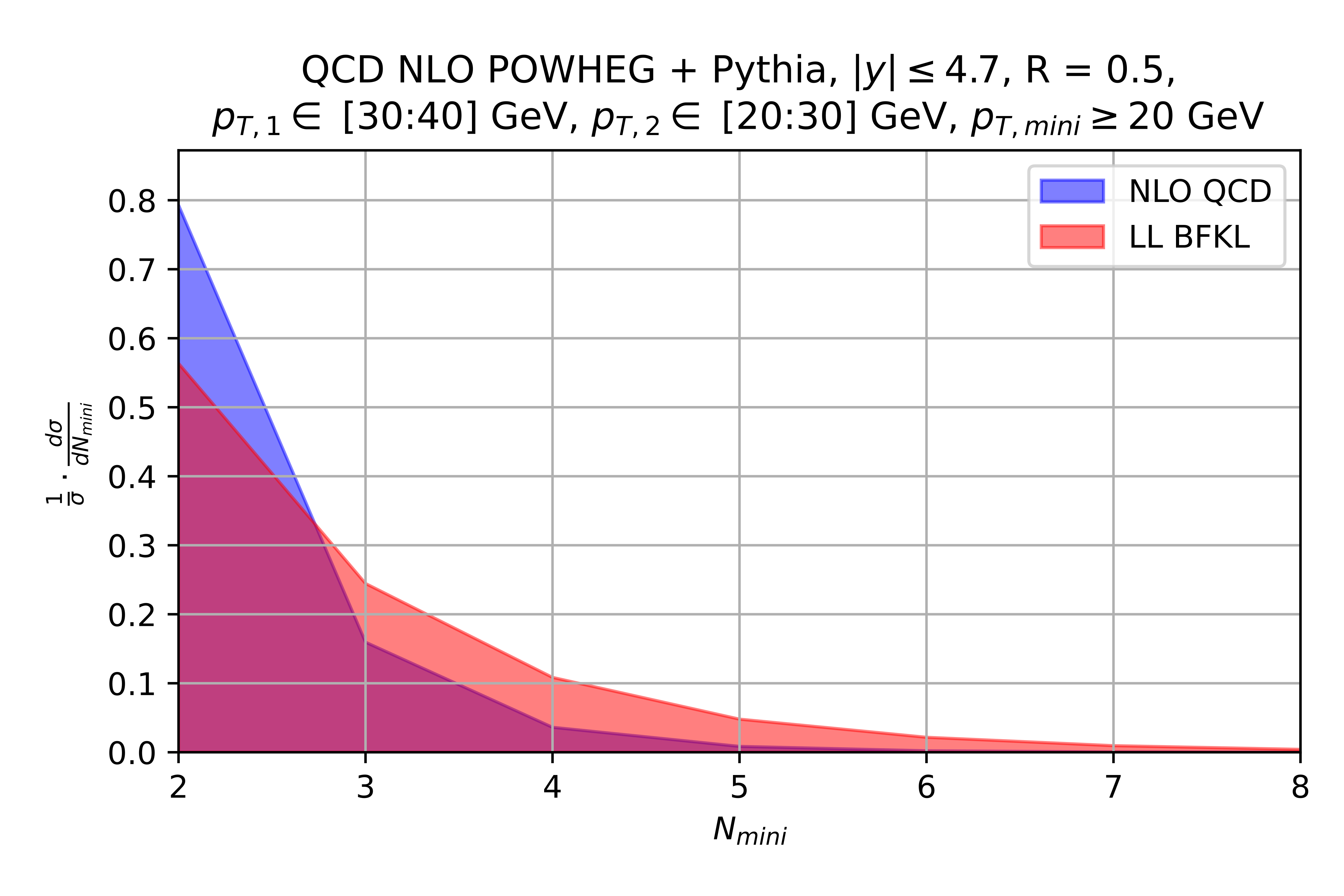}
\caption{Normalized inclusive MN dijet cross section as a function of $\nM$ for $\pM=\SI{10}{GeV}$ (left) and $\pM=\SI{20}{GeV}$ (right) for NLO QCD (blue) and LL BFKL (red).
The events that contribute to these distributions have final state jet multiplicity 4 and above.
}
\label{fig:dsdN}
\end{center}
\end{figure}

A similar reasoning applies to the distributions shown in Fig.~\ref{fig:dsdN}, where $d\sigma/d\nM$ is displayed for both values of $\pM$. The {\tt POWHEG} + {\tt Pythia} approach requires the minijet emission to be mostly collinear in order for more hard jets to be produced. The BFKLex approach does not force any hard ordering on the $\pT$ of the produced minijets. A high $\pT$ minijet can thus be produced at any rapidity of the Regge ladder. Consequently, higher minijet multiplicities for BFKL with respect to NLO QCD are allowed in average as shown in Fig.~\ref{fig:dsdN}. 

\subsection{Rapidity ratio $R_y$}

\begin{figure}
\begin{center}
\includegraphics[width=8.cm]{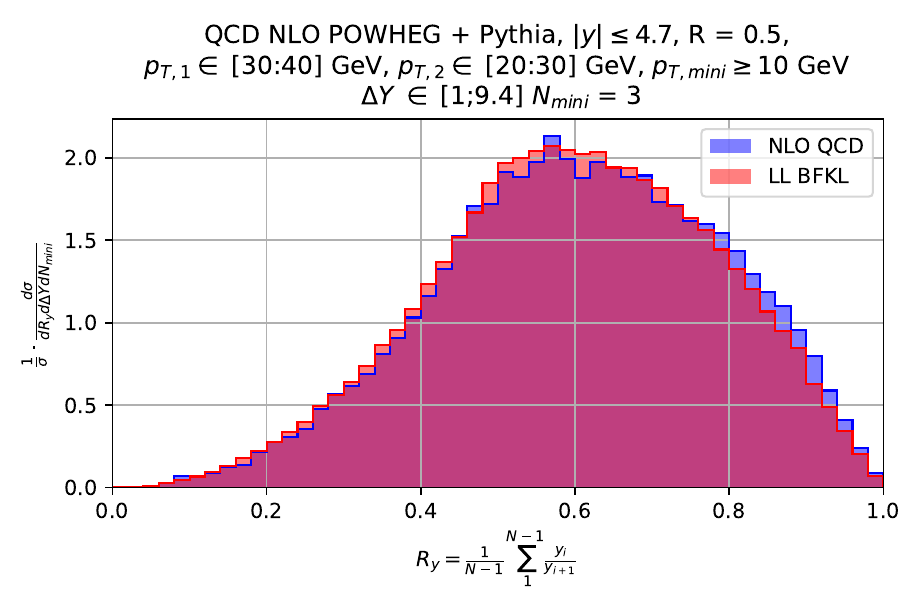}\includegraphics[width=8.cm]{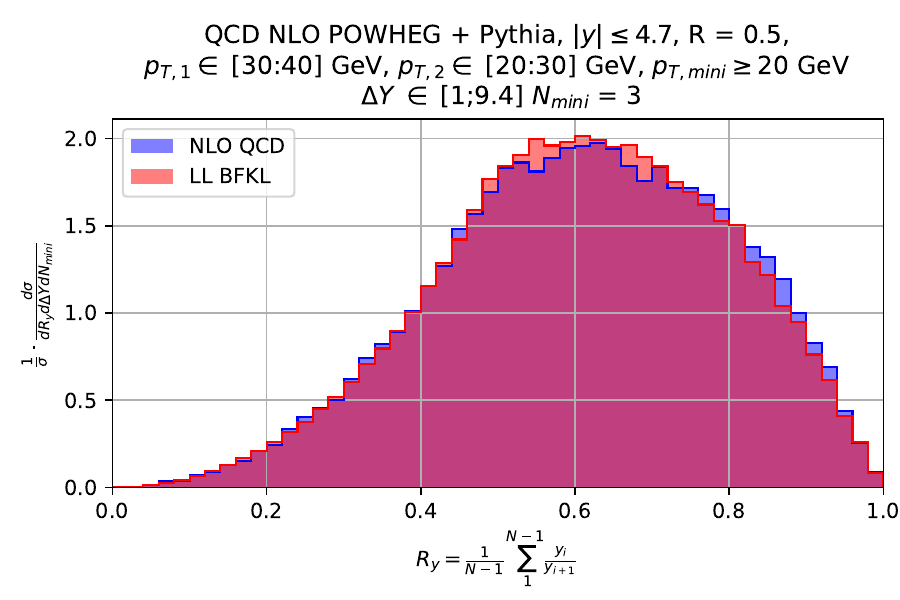}
\includegraphics[width=8.cm]{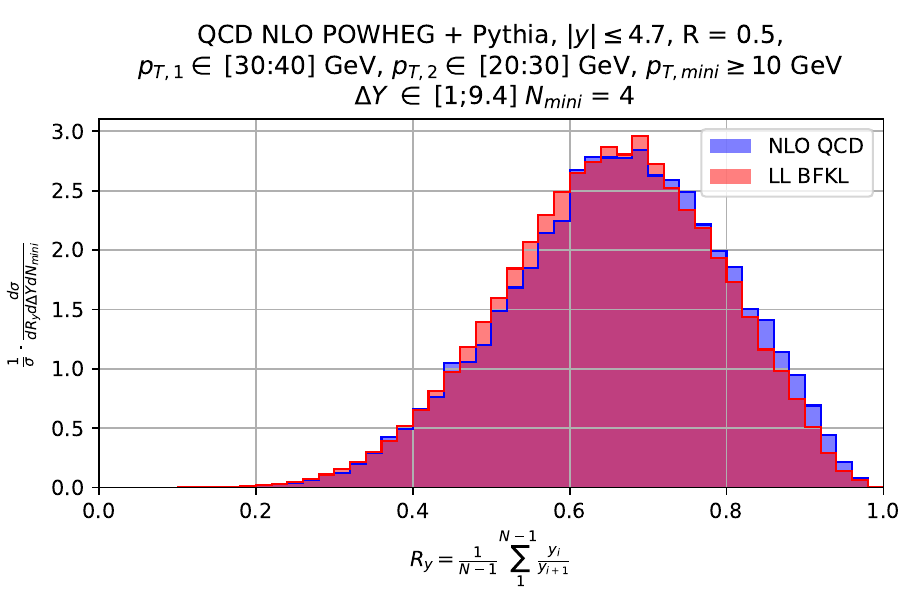}\includegraphics[width=8.cm]{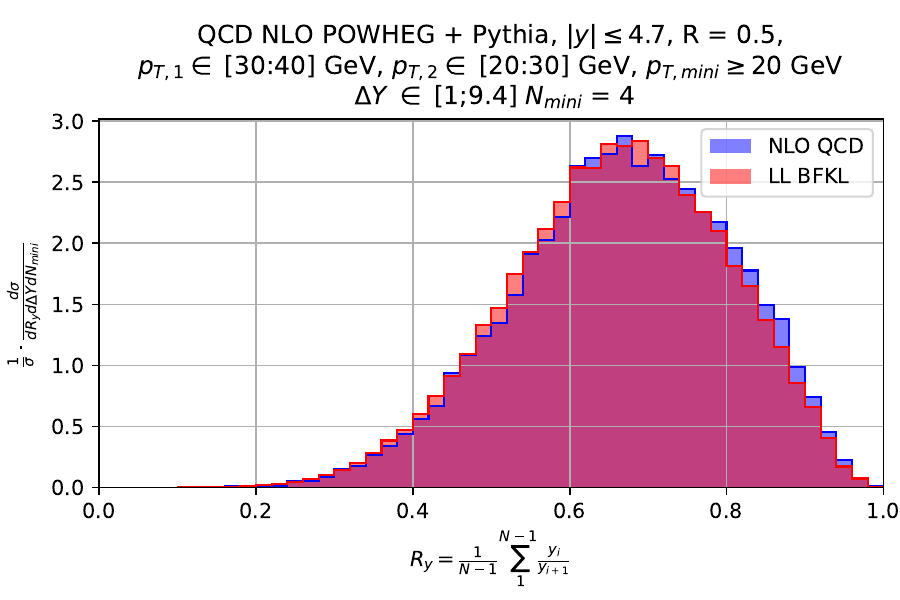}
\includegraphics[width=8.cm]{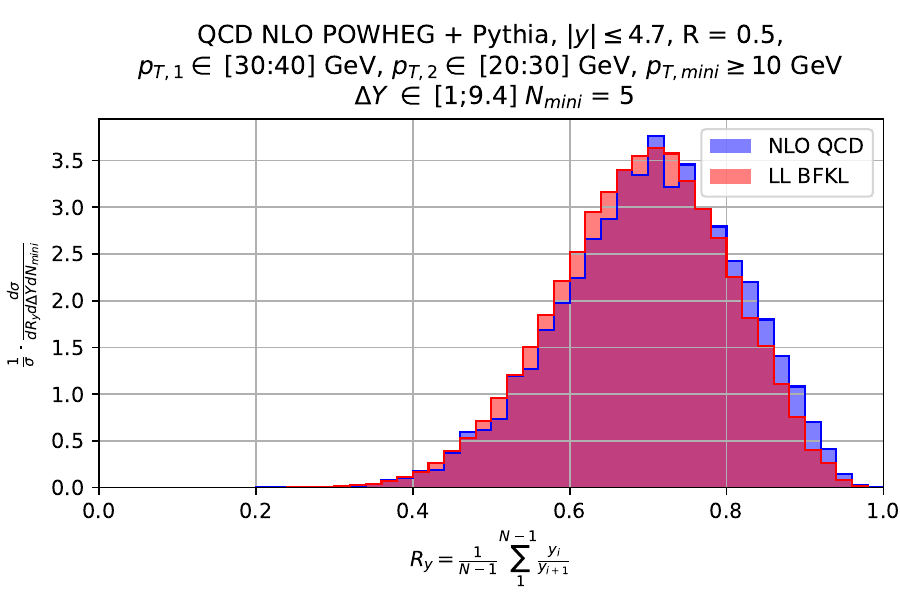}\includegraphics[width=8.cm]{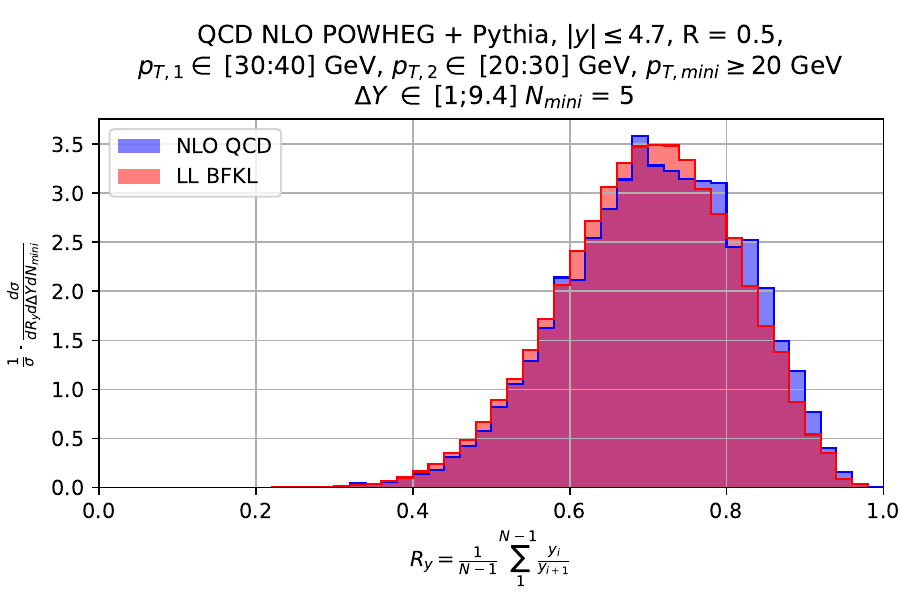}
\caption{$R_y$ for $\pM=\SI{10}{GeV}$ (left column) and $\pM=\SI{20}{GeV}$ (right column), with $\nM=3,4,5$ from top to bottom and $\Delta Y\in[1,9.4]$ for NLO QCD (blue) and LL BFKL (red).}
\label{Ry1to10}
\end{center}
\end{figure}

\begin{figure}
\begin{center}
\includegraphics[width=8.cm]{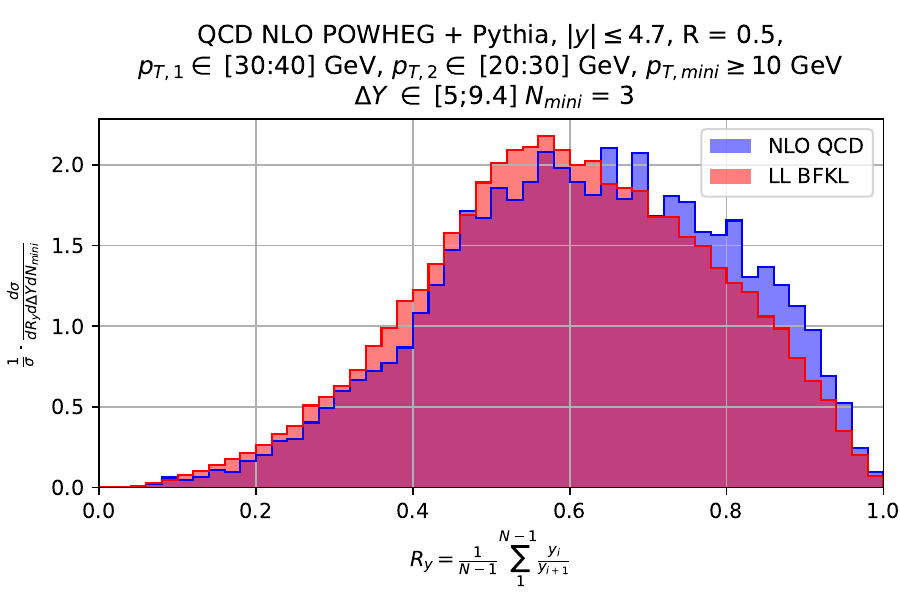}\includegraphics[width=8.cm]{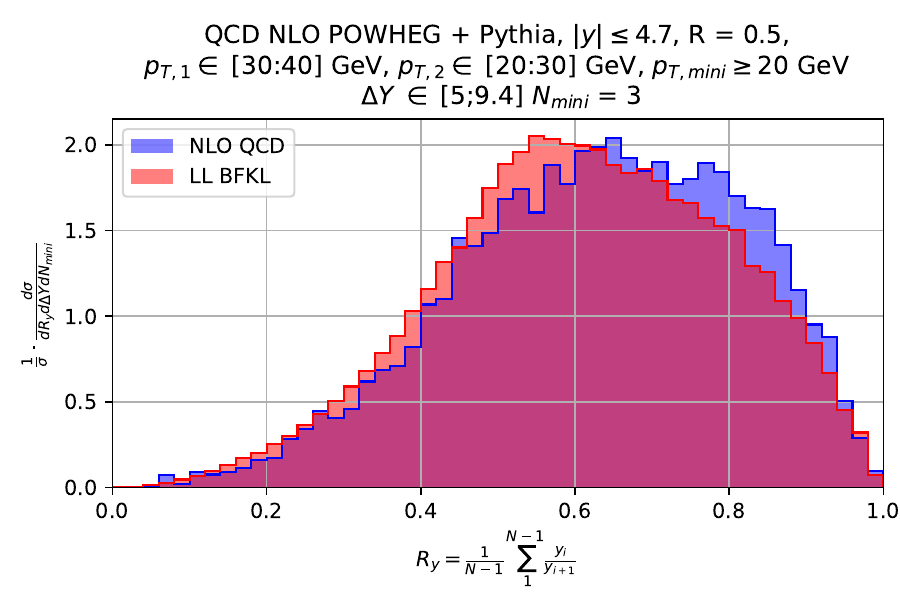}
\includegraphics[width=8.cm]{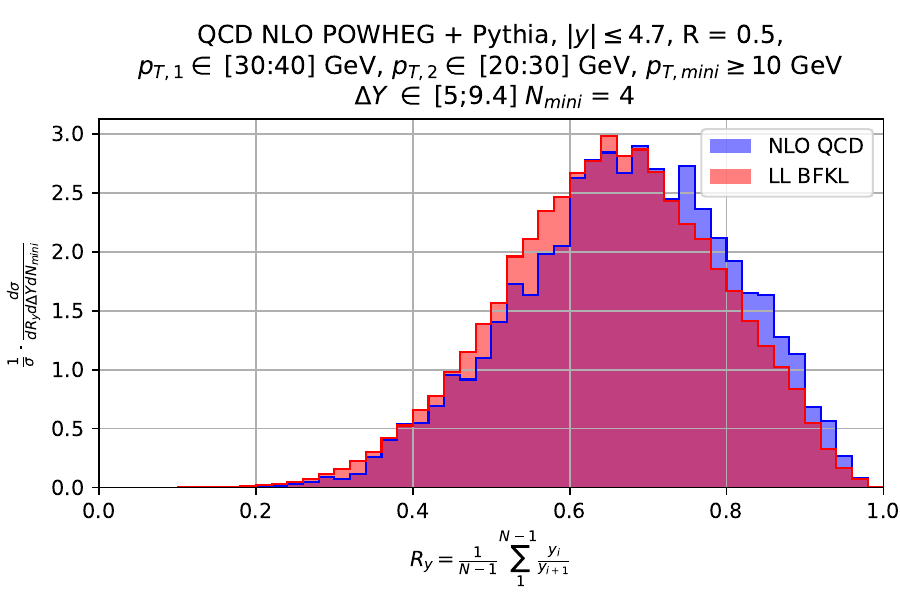}\includegraphics[width=8.cm]{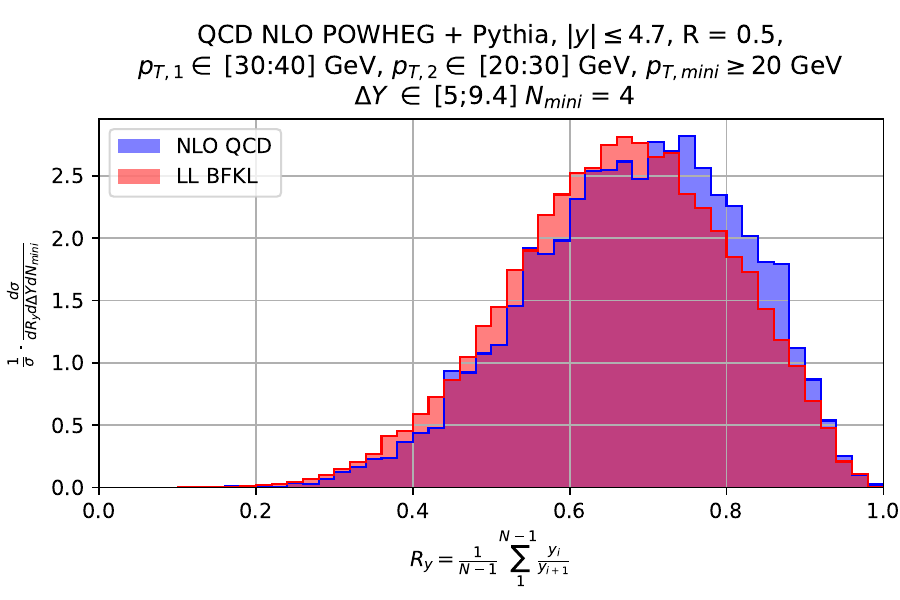}
\includegraphics[width=8.cm]{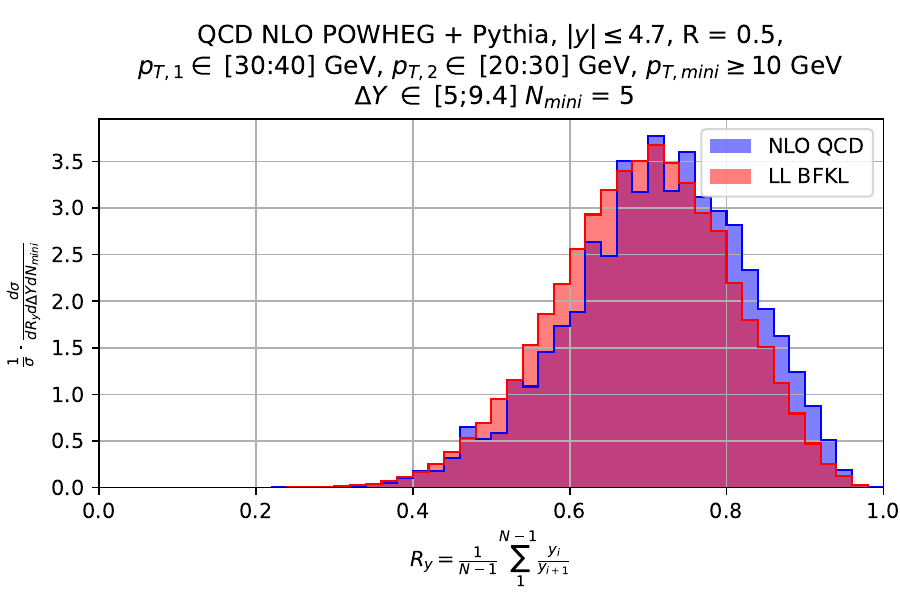}\includegraphics[width=8.cm]{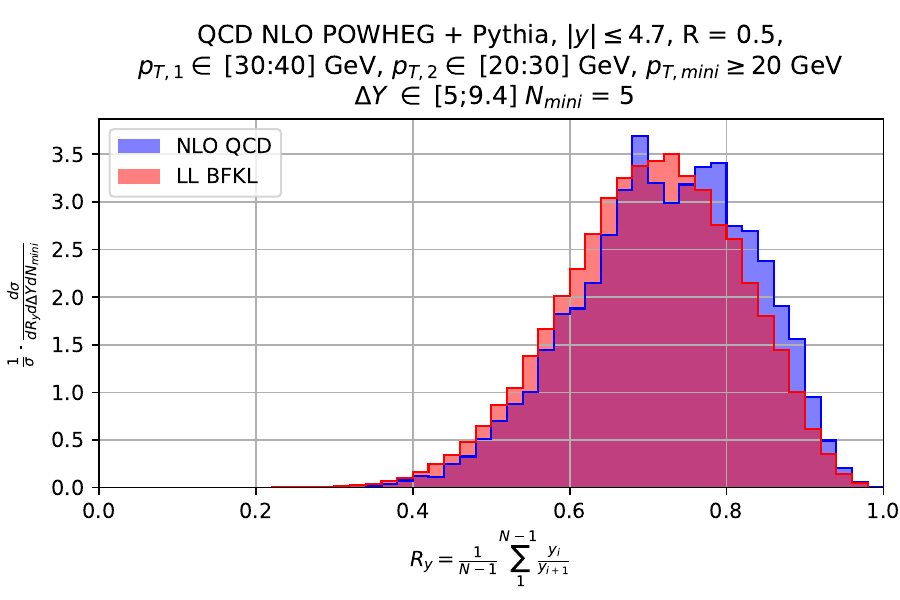}
\caption{$R_y$ for $\pM=\SI{10}{GeV}$ (left column) and $\pM=\SI{20}{GeV}$ (right column), with $\nM=3,4,5$ from top to bottom and $\Delta Y\in[5,9.4]$ for NLO QCD (blue) and LL BFKl (red).}
\label{Ry5to10}
\end{center}
\end{figure}

The rapidity ratio observable is defined by
\begin{equation}
    R_y=\frac{1}{\nM-1}\sum_{i=1}^{\nM-1}\frac{y_i}{y_{i+1}}.
\end{equation}
It is a rather simple expression with values in the interval [0;1]. It is noteworthy that both the {\tt BFKLex} and the NLO fixed-order QCD calculation generate very similar distributions (see Figs.~\ref{Ry1to10} and~\ref{Ry5to10}), considering the quite different physics dynamics implemented by the two approaches. \\

In particular, for the scenario where the multi-Regge kinematics is more applicable, that is, for MN jets with large rapidity separation, we observe that the BFKL prediction is always slightly shifted towards smaller values of $R_y$ than NLO QCD {\tt POWHEG} + {\tt PYTHIA8} (see Fig.~\ref{Ry5to10} for $\Y\in[5;9.4]$), {\it i.e.} BFKL generates smaller $y_i/y_{i+1}$ ratios and tends to have a smaller rapidity interval between successive minijet emissions. This behavior is independent of the final state multiplicity and we find that it is a rather general trend. It is worth noting that the range of possible values for $R_y$ decreases as the final state multiplicity increases. This is because the maximum total $\Y$ is fixed in these plots. For a given multiplicity $\nM$ the mean rapidity separation is $\Y/N$  which decreases as $\nM$ increases leading to narrower distributions. The experimental resolution cutoff $p_{T, {\rm mini}}$ has a minor impact. \\

When we allow for the rapidity difference between the two leading MN jets to take values in the interval $\Y \in [1,9.4]$, see Fig.~\ref{Ry1to10}, we see that the distributions from the two approaches become more similar. It is rather remarkable as it implies that the multi-gluon generator {\tt BFKLex} is not too far from the parton shower contained in \texttt{Pythia}8 in terms of populating with minijets the rapidity space between the bounding MN jets.\\

Shifting from Fig.~\ref{Ry1to10} to Fig.~\ref{Ry5to10}, we notice that, while the NLO QCD prediction appears to be slightly moved to the right compared to the one for $\Y>1$, the BFKL one does not change much. The BFKL distributions, for $\nM$ fixed, remarkably show little difference between the $\Y>1$ and $\Y>5$ cases.  This is a rather clear signal that BFKL favors a more homogenous spread of emissions in the rapidity space. For the event shape under study, $y_n\approx n\cdot\Delta y$, and the separation between the different $y_i$ terms cancels out due to the way $R_y$ is defined. Therefore, in the high-energy limit, $R_y$ becomes independent of the rapidity separation and the emission structure in rapidity of BFKLex does not change much with $\Y$. We also observe that $R_{y,\text{NLO QCD}}>R_{y,\text{BFKL}}$, as is expected from collinear emissions leading to $R_y\sim 1$. 

\begin{figure}
\begin{center}
\includegraphics[width=8.cm]{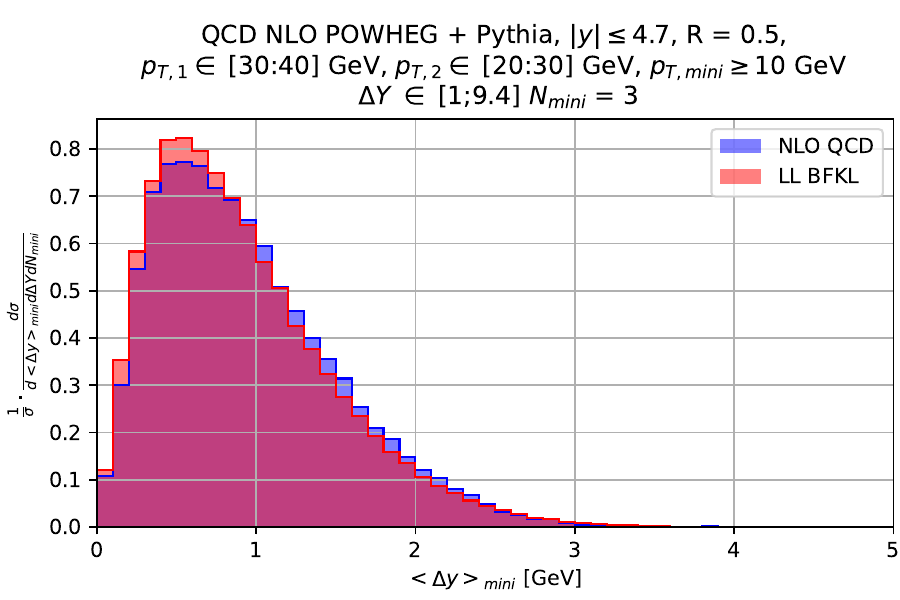}\includegraphics[width=8.cm]{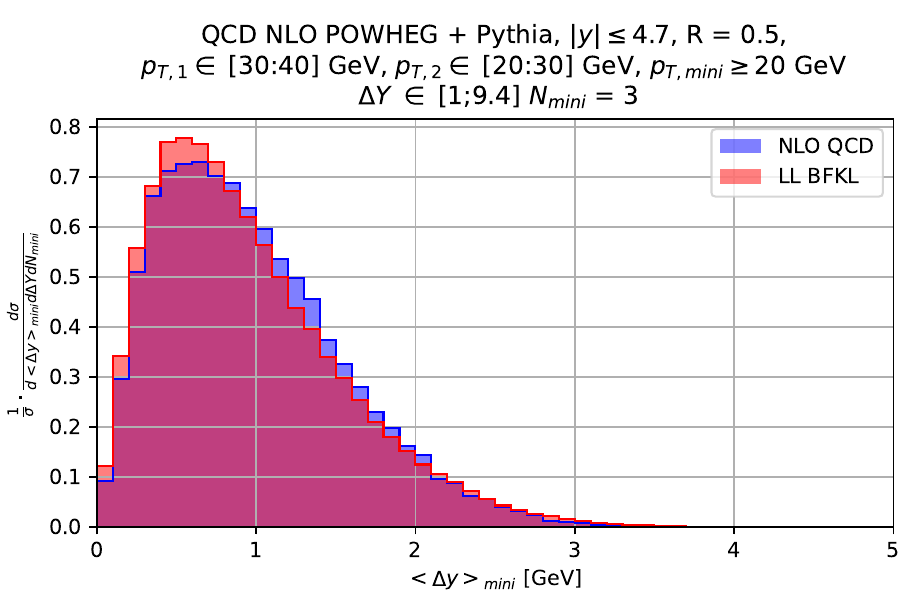}
\includegraphics[width=8.cm]{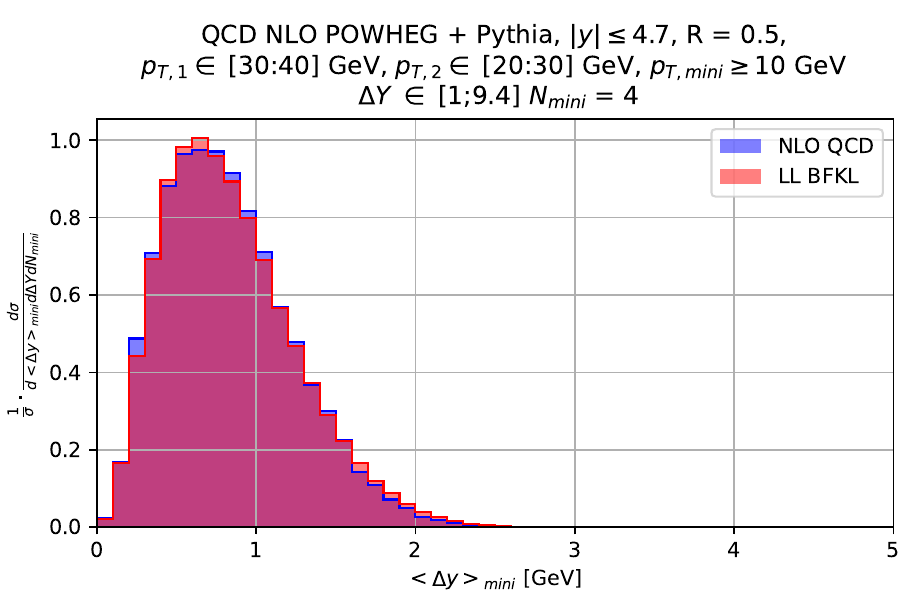}\includegraphics[width=8.cm]{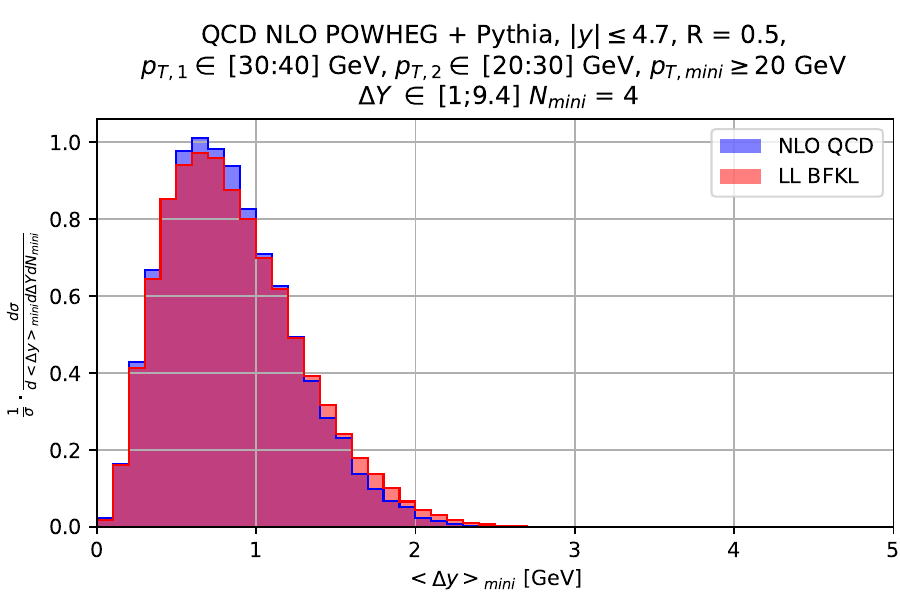}
\includegraphics[width=8.cm]{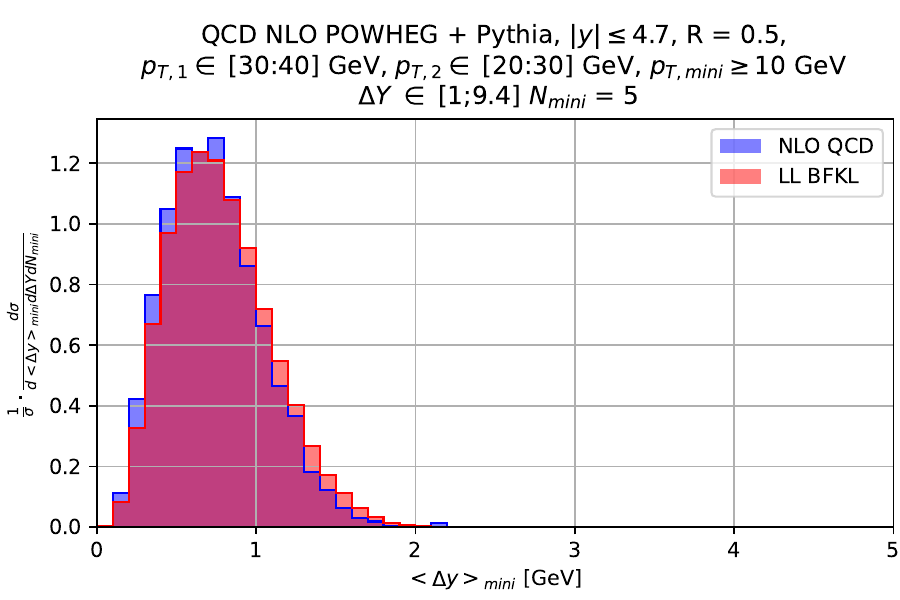}\includegraphics[width=8.cm]{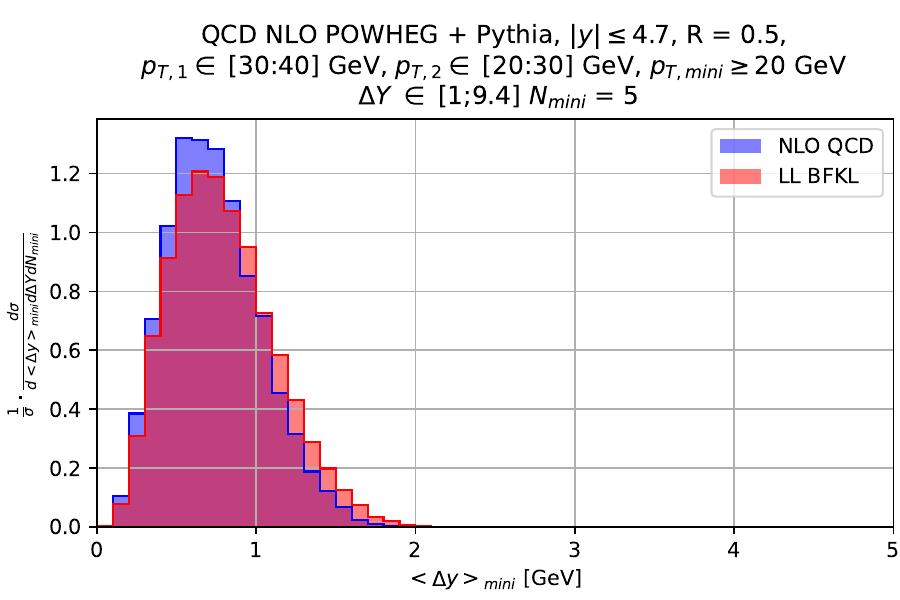}
\caption{$\yobs$ for $\pM=\SI{10}{GeV}$ (left column) and $\pM=\SI{20}{GeV}$ (right column), with $\nM=3,4,5$ from top to bottom and $\Delta Y\in[1,9.4]$ for NLO QCD (blue) and LL BFKL (red).}
\label{DeltaY1to10}
\end{center}
\end{figure}

\begin{figure}
\begin{center}
\includegraphics[width=8.cm]{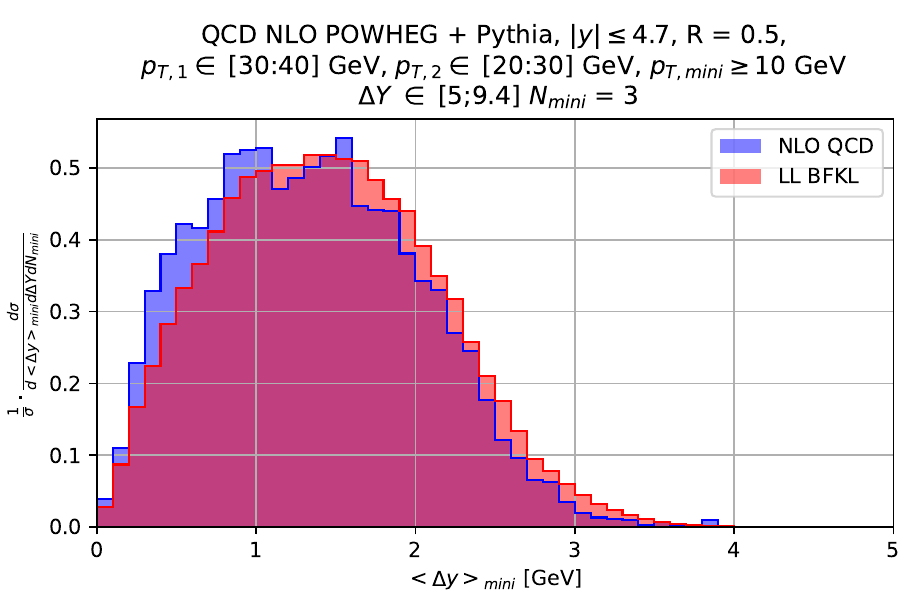}\includegraphics[width=8.cm]{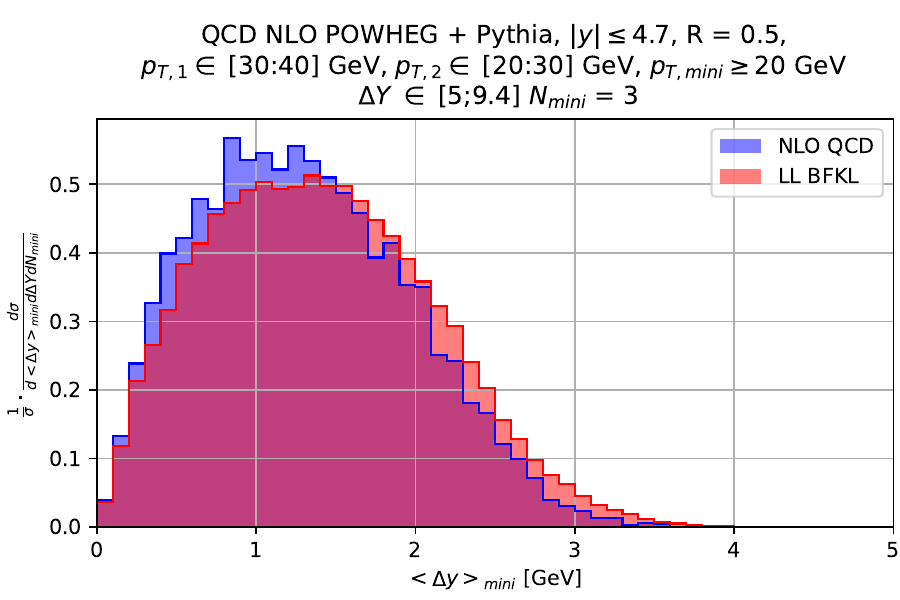}
\includegraphics[width=8.cm]{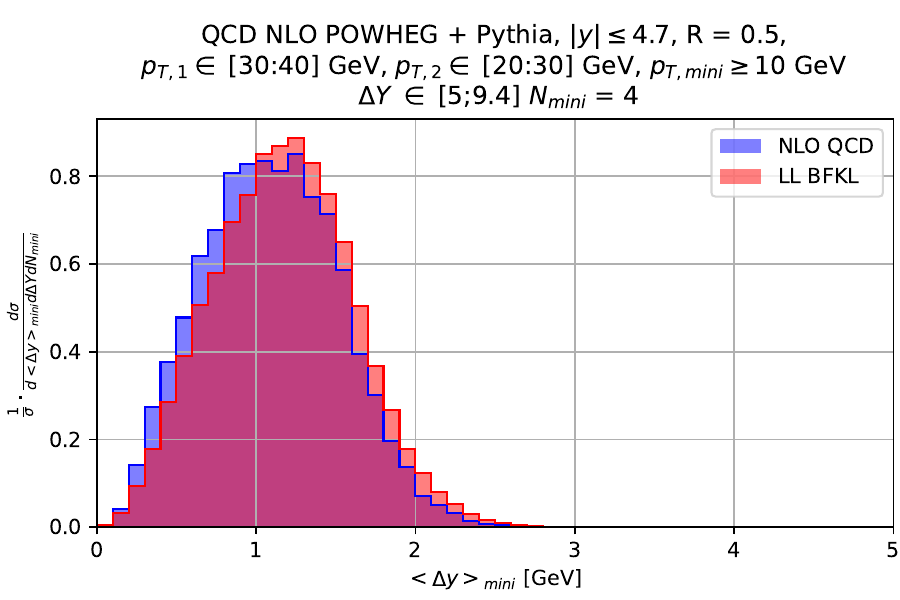}\includegraphics[width=8.cm]{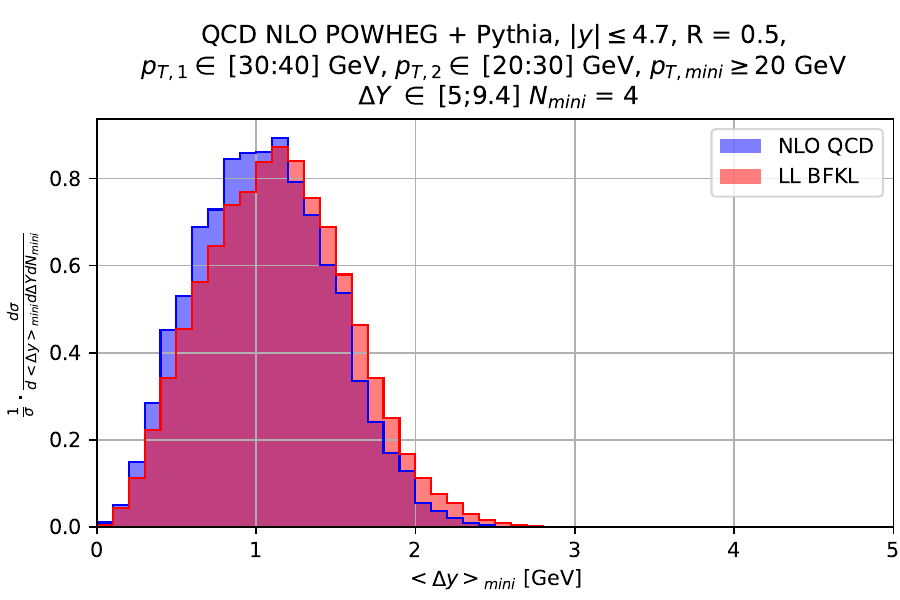}
\includegraphics[width=8.cm]{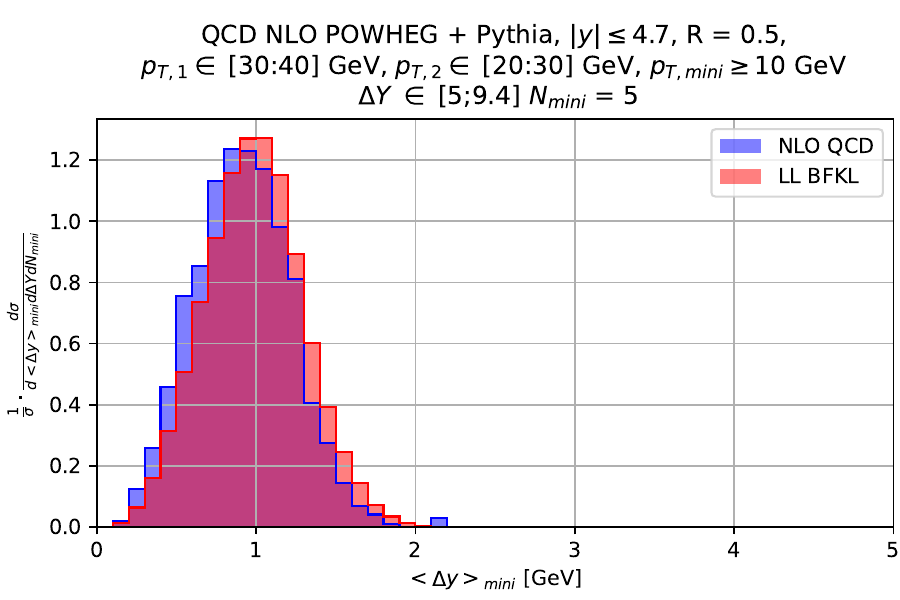}\includegraphics[width=8.cm]{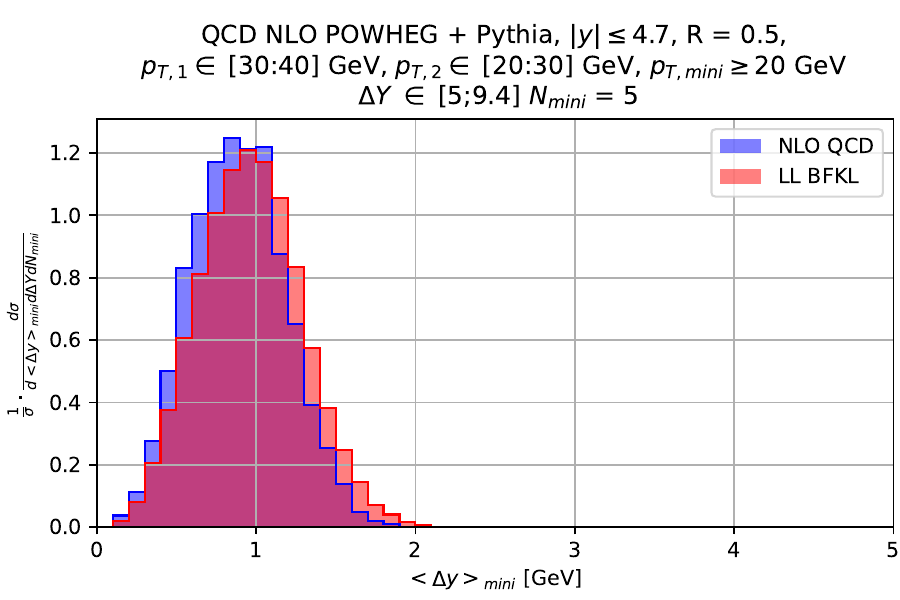}
\caption{$\yobs$ for $\pM=\SI{10}{GeV}$ (left column) and $\pM=\SI{20}{GeV}$ (right column), with $\nM=5,6,7$ from top to bottom and $\Delta Y\in[5,9.4]$ for NLO QCD (blue) and LL BFKL (red).}
\label{DeltaY5to10}
\end{center}
\end{figure}

\subsection{Mean rapidity difference $\yobs$ between minijets}

Similar conclusions can be drawn from the study of the average rapidity difference between successively  emitted (in rapidity) minijets. The related observable to $R_y$ is the average of the rapidity differences between minijets defined as
\begin{equation}
    \yobs=\frac{1}{\nM-1} \sum_{i=1}^{\nM-1}  (y_{i+1}-y_i) \, .  
\end{equation}
Fig. \ref{DeltaY1to10} shows the $\yobs$ distribution with $\Delta Y>1$ for 5,6 and 7 jet events where the number of jets include the two MN jets and the minijets.  It corresponds to the average emission of the minijets in rapidity. There are small differences between the distributions for the two different allowed  values  of $\pM$  and the agreement between the NLO QCD ({\tt POWHEG} + {\tt Pythia}) and {\tt BFKLex} is again striking. \\

\begin{figure}
    \centering
    \includegraphics[width=.70\textwidth]{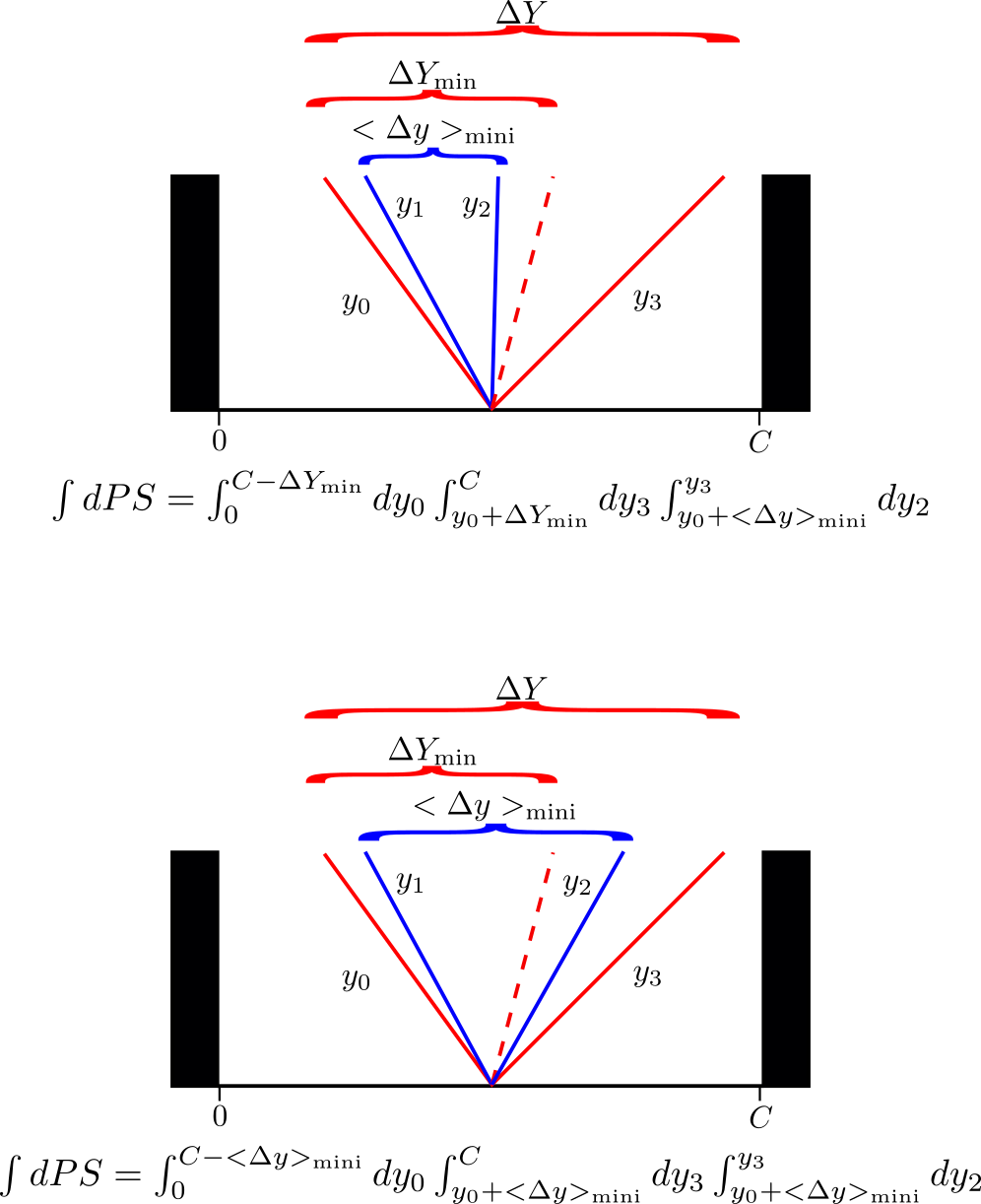}
    \caption{Changes in phase space integral depending on minimum MN jet rapidity separation cut $\Ym$.}
    \label{fig:psint}
\end{figure}

We notice in Fig.~\ref{DeltaY1to10} that the distributions are narrower for a higher jet multiplicity, which is expected because of the  $(\nM-1)^{-1}$ factor in the definition of $\yobs$. The distribution at high $\yobs$ is suppressed since only high-$\Y$ events contribute to large values of $\yobs$.  The agreement of {\tt BFKLex} and {\tt POWHEG}+{\tt Pythia} outside of the BFKL regime at small $\Y$ is noteworthy. We observe small changes in the shapes between $\pM=\SI{10}{GeV}$  and $\pM=\SI{20}{GeV}$. This is a sign that the small $\Y$ domain is in a safe kinematical region, away from soft singularities that would modify the emission patterns in NLO QCD.  We note that $\yobs$ slightly grows with $\pM$ since a higher $\pM$ requirement requires the existence of harder jets, which statistically need to be even more collinear to their parent. If the emissions stem from the tagged MN jets, as it is statistically the case for low-$\Y$ events, $\yobs$ would increase minimally. \\

In Fig. \ref{DeltaY5to10} we show the distributions of $\yobs$ for $\Delta Y>5$. For fixed $\nM$, the distributions are now wider and peak at higher $\yobs$. The reason for this is shown in Fig.~\ref{fig:psint}: due to the available phase space, $d\sigma/d\yobs$ grows with two less powers of $\yobs$ at $\yobs\,<\Delta Y_\text{min}/(\nM-1)$ than at $\yobs\,>\Delta Y_\text{min}/(\nM-1)$. Increasing the $\Delta Y_\text{min}$ value induces a less steep growth on the left part of the curve. In the high-$\Y$ case, the peak value now decreases with $\pM$ for low $\nM$. In NLO QCD this is because minijets are then more likely to be emitted from a central hard jet whereas for large $\nM$  it is more probable that emissions from the MN jets will give the outermost minijets. \\

It is again remarkable that the distributions from the BFKL and NLO QCD calculations are so similar in a kinematical domain that is a priori sensitive to BFKL resummation effects. It shows that these observables are more sensitive to multi-gluon emissions between the two MN jets which seems to be a general property of QCD. The experimental measurement of these minijet emissions and the comparison with the theoretical predictions should be of great interest for future works.

 \subsection{Invariant mass ratio $R_{ky}$}

In order to enhance the differences between the distributions from the high-energy resummation and the fixed-order calculation, we allow the rapidity ratios to acquire a transverse-momentum dependence. To this end, we define the observable
\begin{eqnarray}
R_{ky} &=& \frac{1}{\nM-1} \sum_{i=1}^{\nM-1} \frac{k_i e^{y_i}}{k_{i+1} e^{y_{i+1}}} \, ,
\end{eqnarray}
which is related to the average ratio of the invariant mass of successively emitted minijet pairs. This variable is less restricted than $R_y$, since the presence of the transverse momentum allows for more scales to have an impact on the distribution, and it can take values greater than one. \\

\begin{figure}
\begin{center}
\includegraphics[width=8.cm]{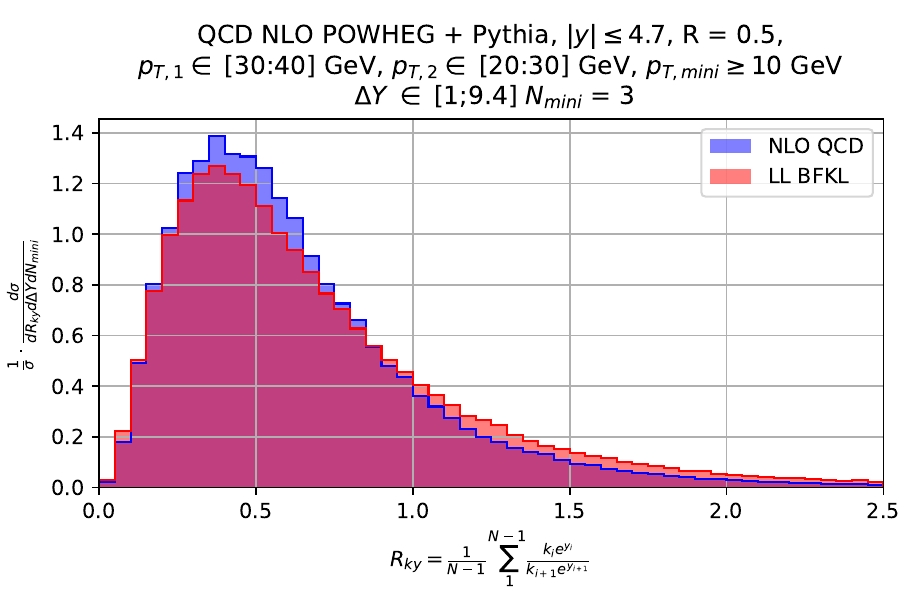}\includegraphics[width=8.cm]{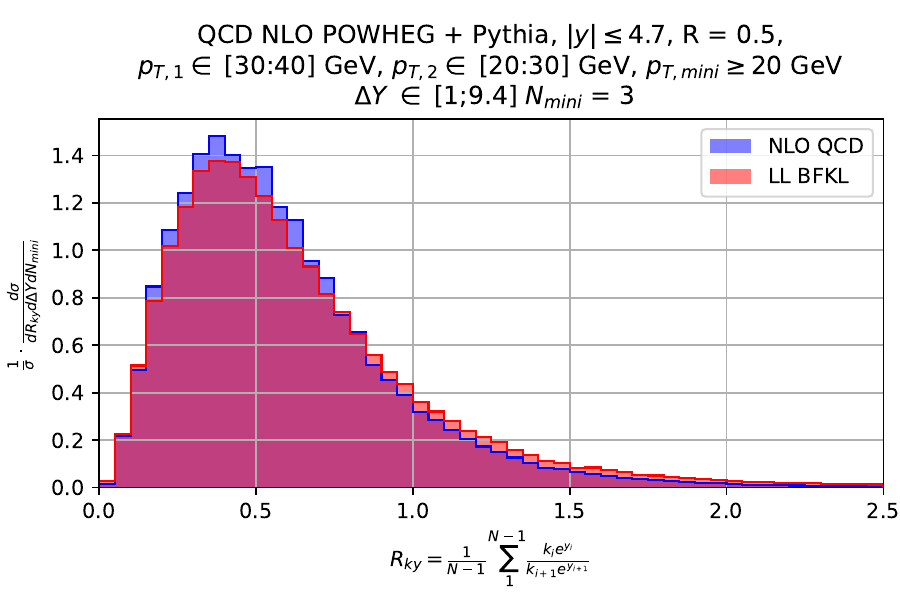}
\includegraphics[width=8.cm]{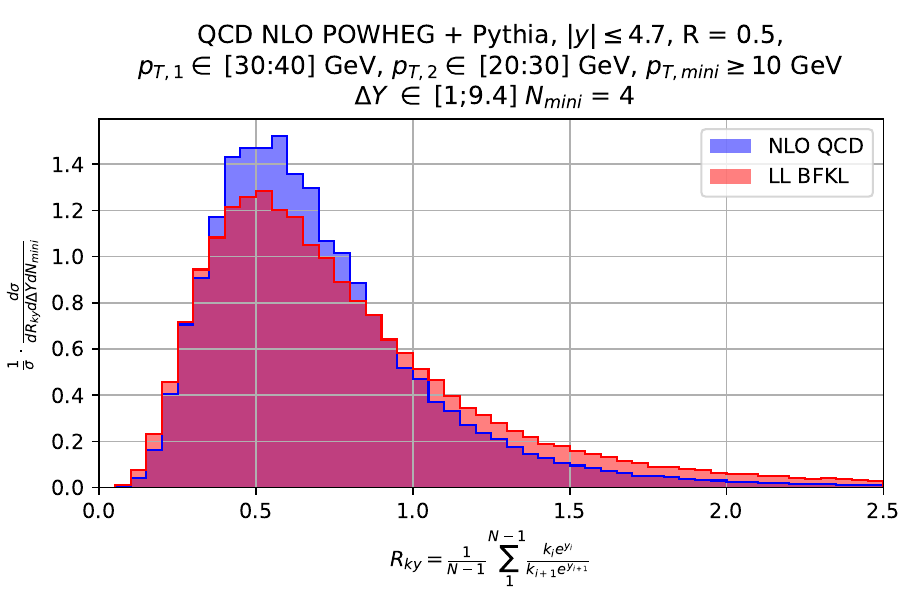}\includegraphics[width=8.cm]{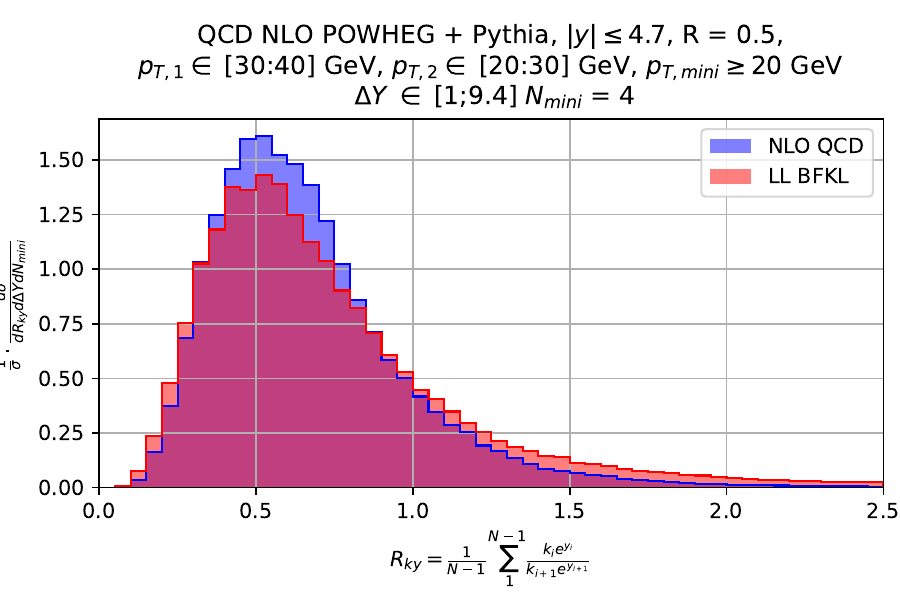}
\includegraphics[width=8.cm]{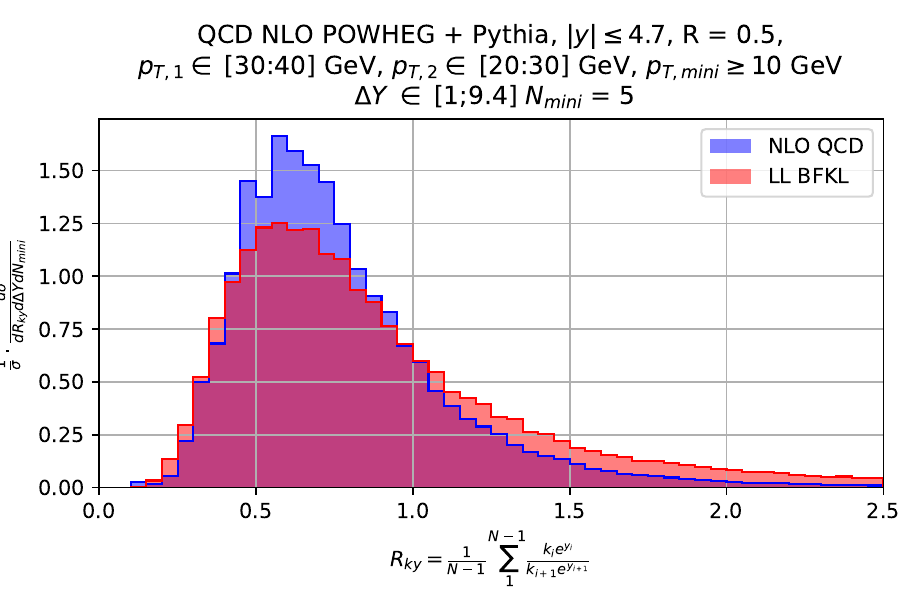}\includegraphics[width=8.cm]{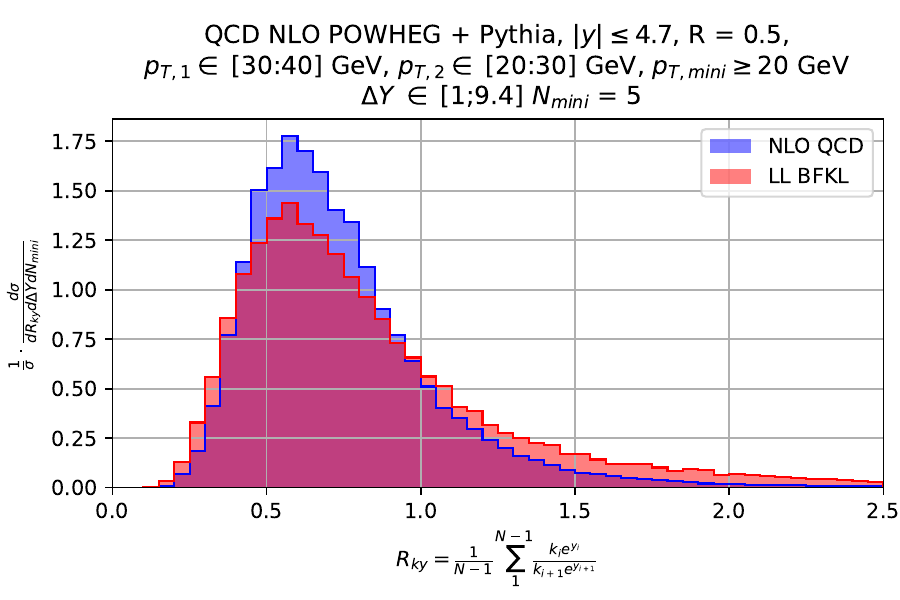}
\caption{$R_{ky}$ for $\pM=\SI{10}{GeV}$ (left column) and $\pM=\SI{20}{GeV}$ (right column), with $\nM=3,4,5$ from top to bottom and $\Delta Y\in[1,9.4]$ for NLO QCD (blue) and LL BFKl (red).}
\label{Rky1to10}
\end{center}
\end{figure}

\begin{figure}
\begin{center}
\includegraphics[width=8.cm]{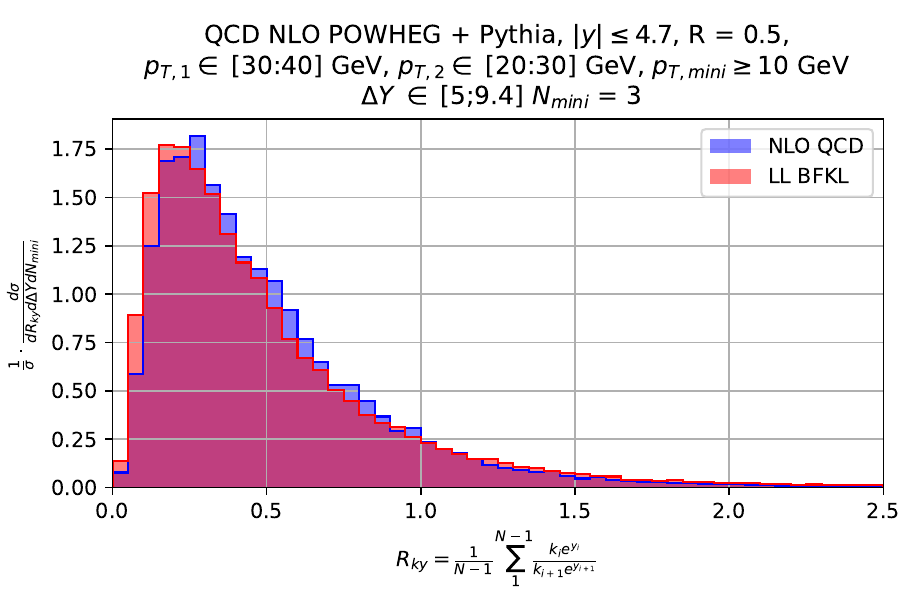}\includegraphics[width=8.cm]{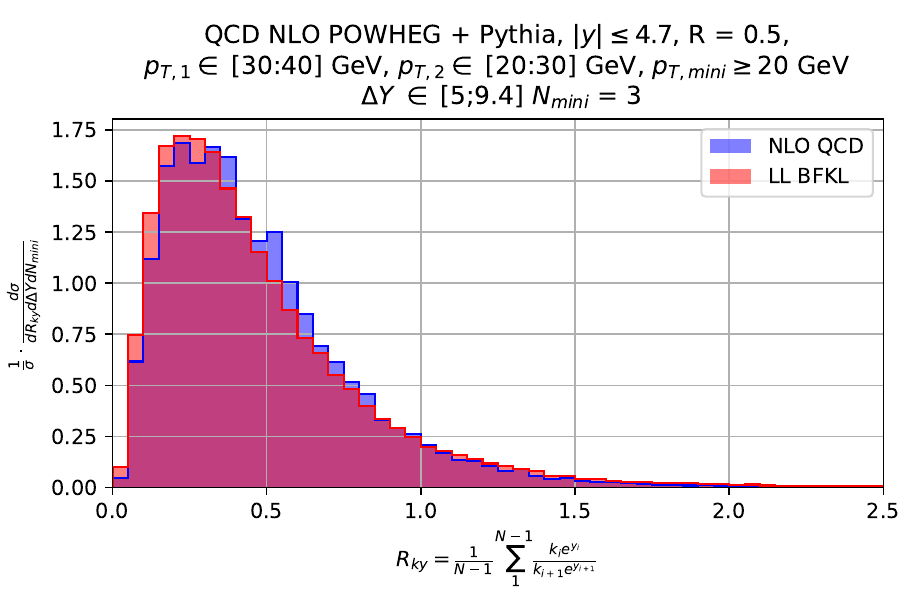}
\includegraphics[width=8.cm]{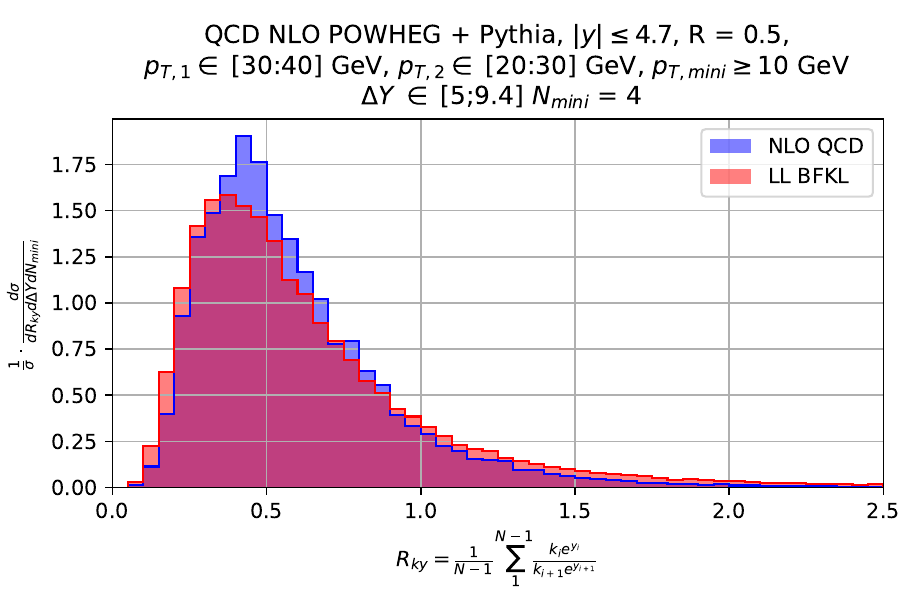}\includegraphics[width=8.cm]{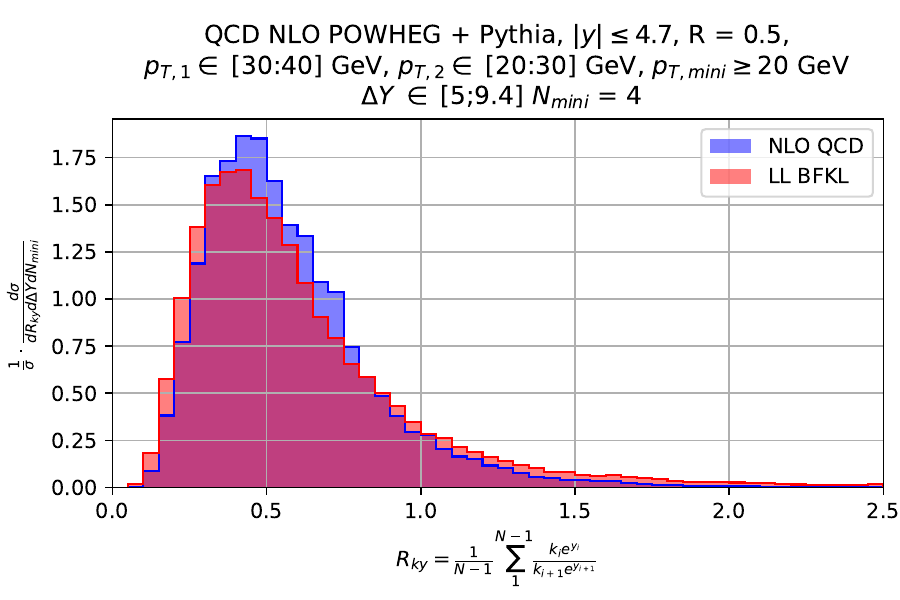}
\includegraphics[width=8.cm]{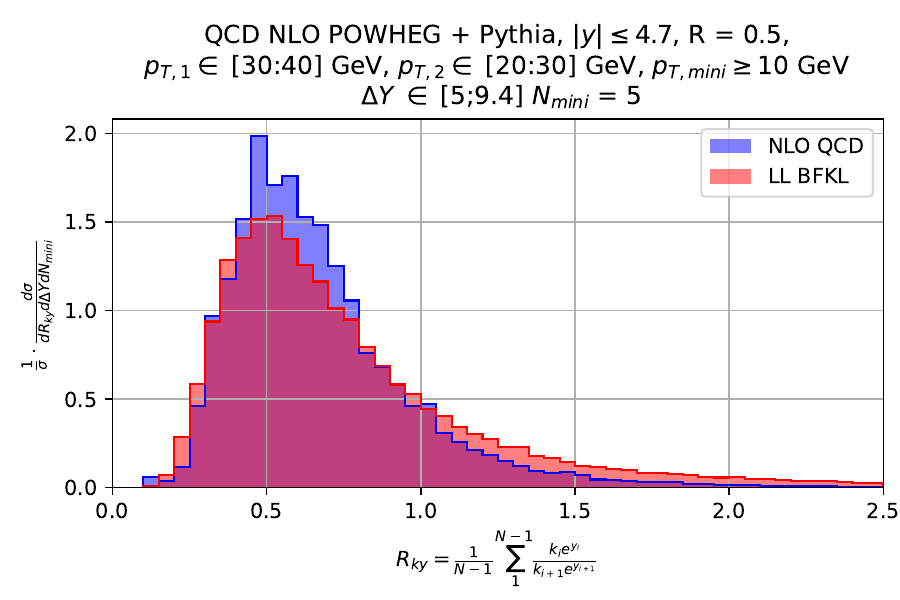}\includegraphics[width=8.cm]{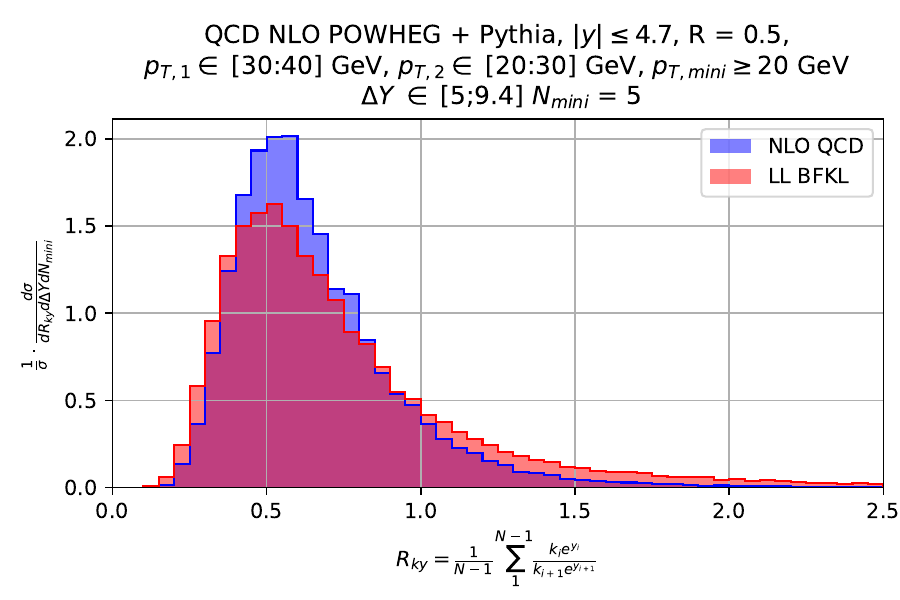}
\caption{$R_{ky}$ for $\pM=\SI{10}{GeV}$ (left column) and $\pM=\SI{20}{GeV}$ (right column), with $N=3,4,5$ from top to bottom and $\Y\in[5,9.4]$ for NLO+PS (blue) and LL BFKL (red).}
\label{Rky5to10}
\end{center}
\end{figure}

In Fig.~\ref{Rky1to10}, the predictions based on these two frameworks yield very similar results, as expected in a region dominated by low-$\Delta Y$ events, cf. Fig.~\ref{dSigmadYfull}.
The distributions for {\tt BFKLex} and {\tt POWHEG} + {\tt Pythia8} peak at very similar values of $R_{ky}$.
The width of the distributions remains similar as the minijet multiplicity $\nM$ increases. The behavior at the higher-tails of the distributions are modestly modified depending on $\nM$. This feature is independent of the cutoff $\pT$ value used for the minijets and also independent from either looking at the full allowed set of rapidity differences between the MN jets $\Y \in [1,9.4]$ or only for $\Y \in [5,9.4]$ (see Fig.~\ref{Rky5to10}). This is because the only possible contribution to $R_{ky}\approx 0$ originates from events with $|y_{i+1}-y_i|$ large for all successive minijet pairs, which is impossible at small enough $\Y/\nM$. Hence, these features are not only characteristic of the high-energy region, but also of the full range in phase space. \\

The width of the distributions from {\tt BFKLex} is wider, most probably due to the fact that no ordering in $\pT$ is forced on the minijet emissions. $R_{ky}$ is sharper for the NLO QCD result, and the effect increases with $\nM$. The events generated in {\tt POWHEG} + {\tt Pythia}, especially at high $\nM$, are dominated by a given number of showered jets. Those have tighter kinematical restrictions than in {\tt BFKLex}, so their distribution is narrower. 

The observed $R_{ky,\text{NLO QCD}}>R_{ky,\text{BFKL}}$ is expected, since the rapidity contribution to $R_{ky}$ becomes minimal for evenly spaced rapidities, $\left.\pdv{R_{ky}}{y_i}\right\vert_{y_i \ll y_{i+1}}=0$.

We conclude the section by emphasizing the need of experimental measurements with larger minijet multiplicities and a lower jet $\pT$ scale in order to investigate properly the regions of phase space where BFKL effects are expected to be more distinct from the NLO+PS approach.

\subsection{Average minijet transverse momentum $\langle \pT \rangle_{\text{mini}}$ }

The differences observed between the $R_{ky}$ distributions from BFKL and NLO QCD after increasing the minijet multiplicity due to the introduced $\pT$ dependence motivate us to study in this subsection  the average $\pT$ of the tagged minijets, that is, $\langle \pT \rangle_{\text{mini}}$. \\

\begin{figure}
\begin{center}
\includegraphics[width=8.cm]{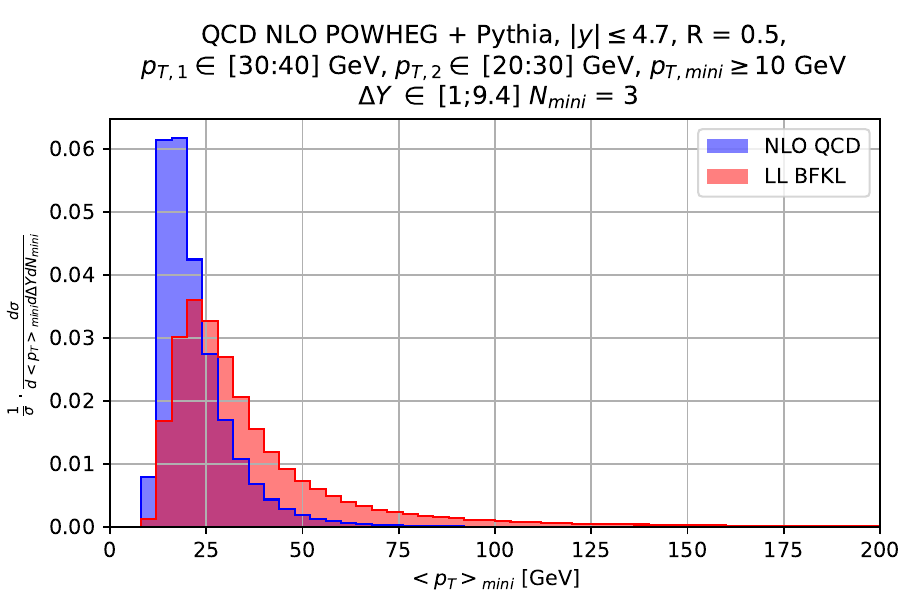}\includegraphics[width=8.cm]{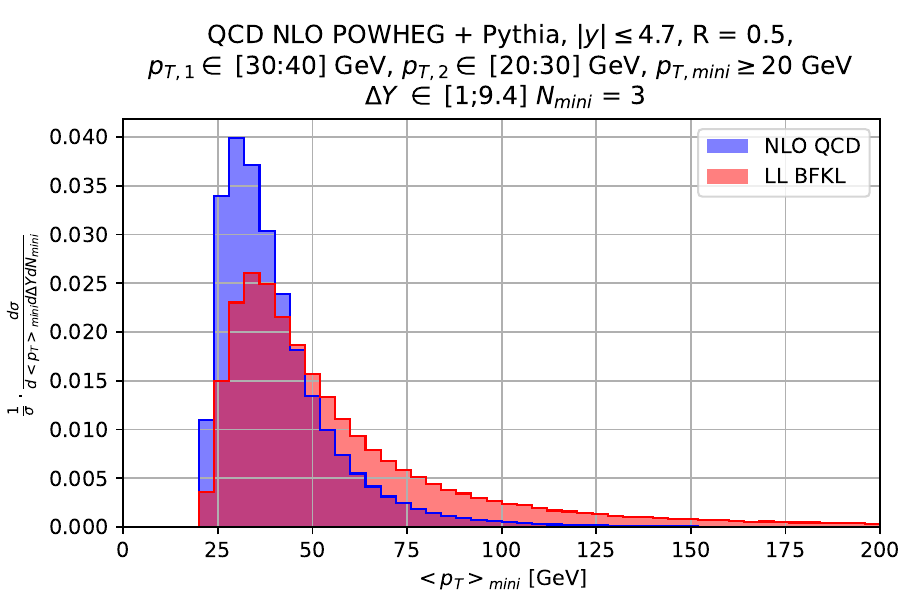}
\includegraphics[width=8.cm]{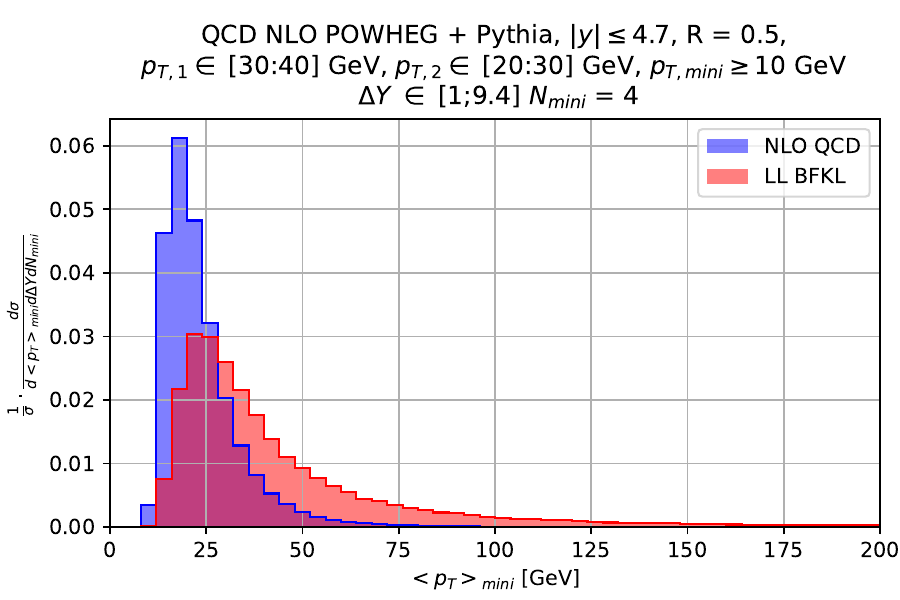}\includegraphics[width=8.cm]{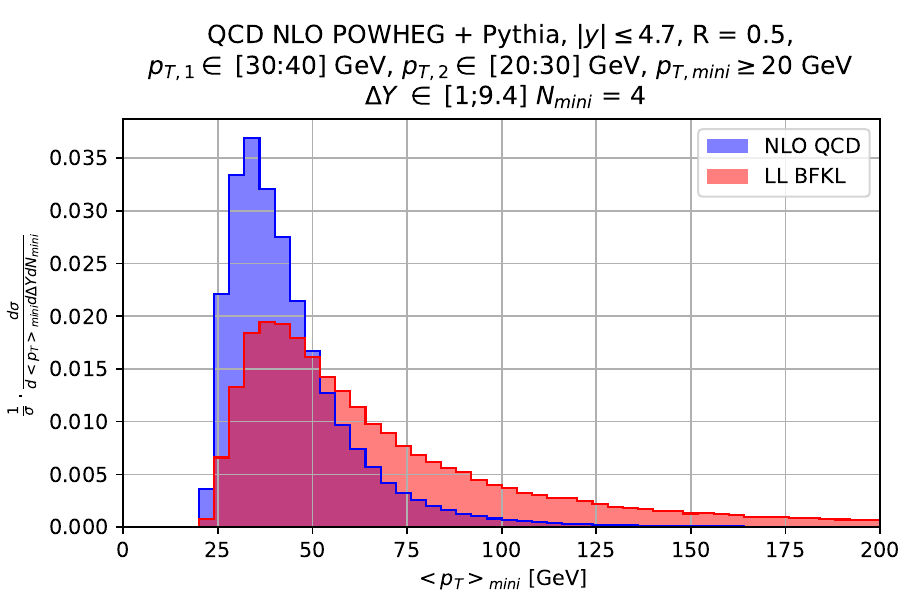}
\includegraphics[width=8.cm]{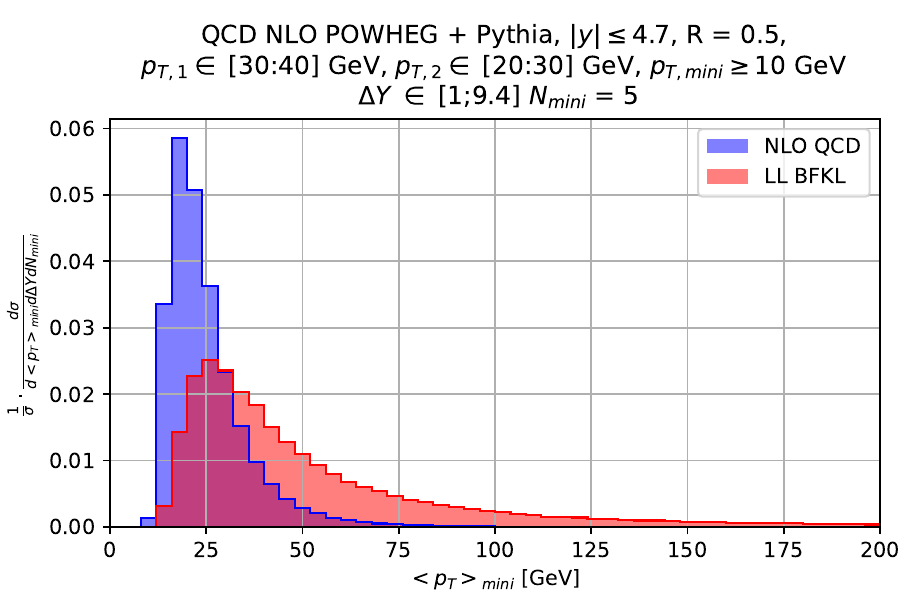}\includegraphics[width=8.cm]{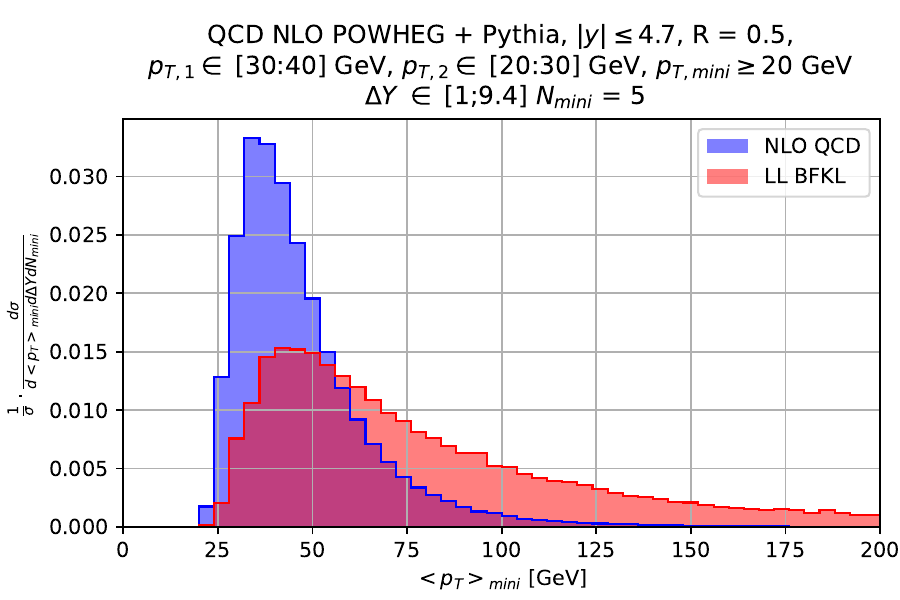}
\caption{$\pobs$ for $\pM=\SI{10}{GeV}$ (left column) and $\pM=\SI{20}{GeV}$ (right column), with $\nM=3,4,5$ from top to bottom and $\Delta Y\in[1,9.4]$ for NLO QCD (blue) and LL BFKL (red).}
\label{AvP1to10}
\end{center}
\end{figure}

\begin{figure}
\begin{center}
\includegraphics[width=8.cm]{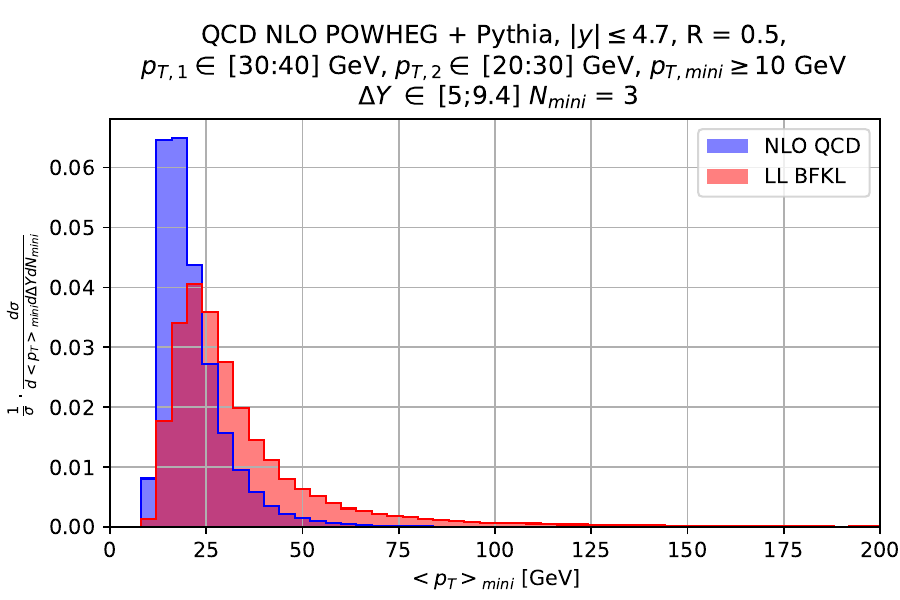}\includegraphics[width=8.cm]{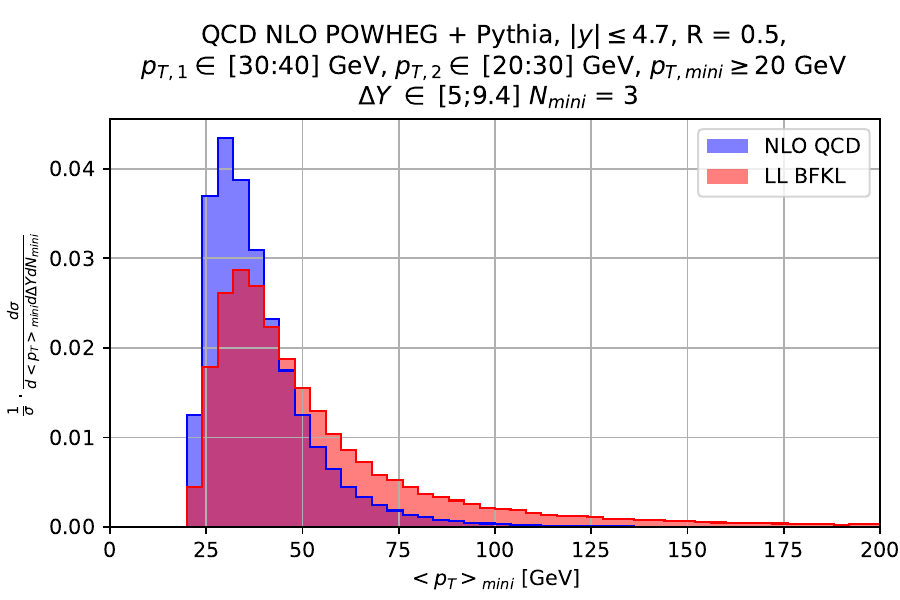}
\includegraphics[width=8.cm]{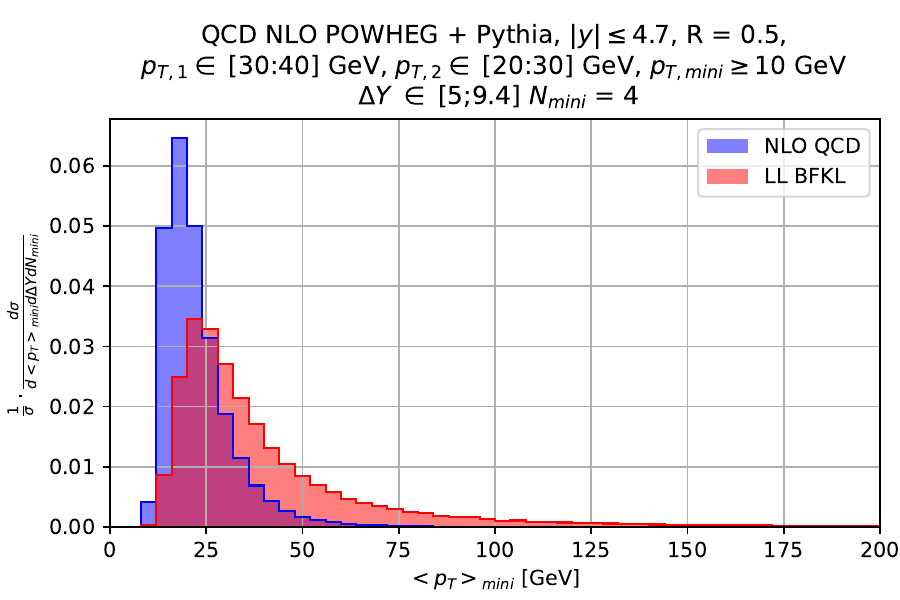}\includegraphics[width=8.cm]{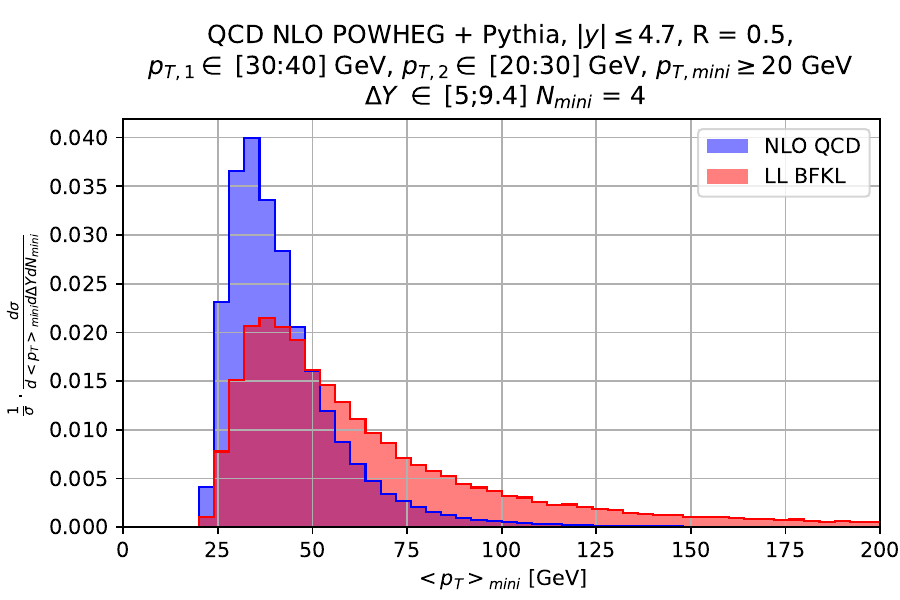}
\includegraphics[width=8.cm]{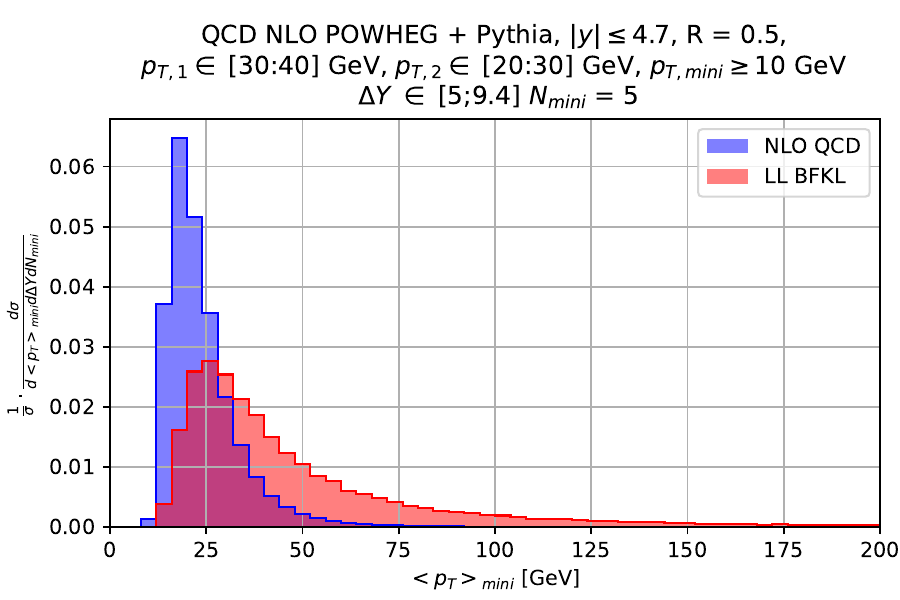}\includegraphics[width=8.cm]{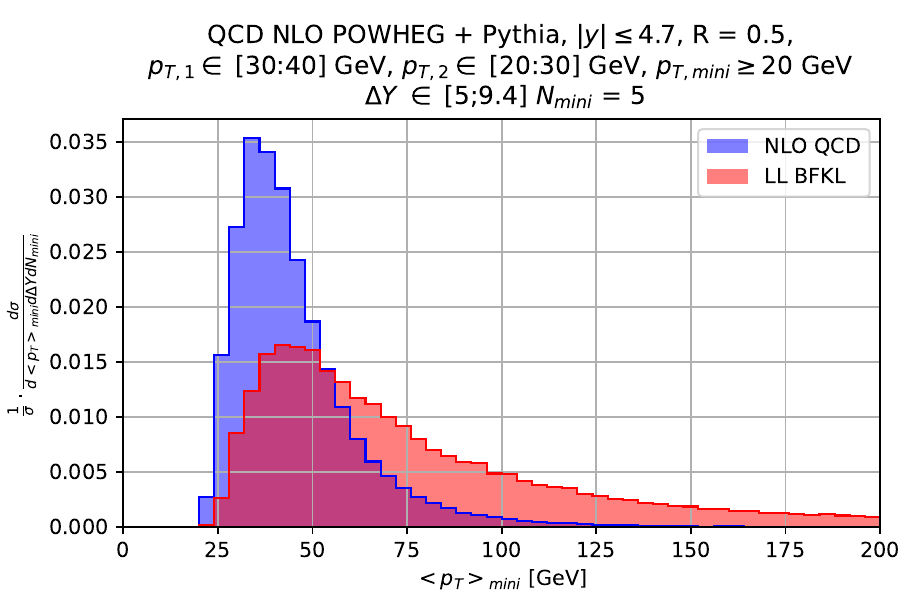}
\caption{$\pobs$ for $\pM=\SI{10}{GeV}$ (left column) and $\pM=\SI{20}{GeV}$ (right column), with $\nM=3,4,5$ from top to bottom and $\Y\in[5,9.4]$ for NLO QCD (blue) and LL BFKL (red).}
\label{AvP5to10}
\end{center}
\end{figure}

We present our results in Figs.~\ref{AvP1to10} and \ref{AvP5to10}. The NLO+PS approach predicts a higher peak (and obviously a broader overall distribution) for lower values of $\pT$ than ${\tt BFKLex}$.
The difference between the two approaches is greater with larger minijet multiplicity $\nM$. The shape of the curves does not change significantly when we change $\Y \in [1,9.4]$ to  $\Y \in [5,9.4]$, suggesting that they are approximately independent of the rapidity separation of the MN jets. The average $\pT$ is larger as we increase the lower cutoff for the minijets from 10 to 20 GeV in both Figs.~\ref{AvP1to10},~\ref{AvP5to10}, as expected. The peaks of the distributions in both approaches shift slightly to larger values as the minijet multiplicity increases.\\

It is possible that the observed differences in the shapes of the distributions are due to the fact that the NLO+PS event generation enforces 4-momentum conservation, 
whereas the BFKL approach only enforces it in a two-dimensional plane transverse to the beam. Thus, a comparison with the experimental measurements would be instructive to assess the validity of the BFKL effective theory in these preasymptotic energy regimes. In this work, we generate the BFKL distributions by strictly adhering to the original formulation of the effective theory. The inclusion of parton shower effects would tend to lower the average jet hardness, as it is well-known in the literature and in a future work, it would be interesting to see how it would change the \texttt{BFKLex} distributions.\\

We point out that we are effectively working with a ``reduced'' version of BFKL dynamics, which is mainly due to two reasons. First, the momentum scale $p_{T,{\rm min}}$ that we use to tag the minijets reduces a significant portion of the BFKL cross section. For fully-inclusive observables, such as structure functions at small Bjorken $x$ in deep-inelastic scattering, the BFKL Green's function includes large contributions from soft emissions, despite being an infrarred-finite resummation. These contributions are absent in our analysis, hence we are only probing the high-$\pT$ regime of BFKL. In this regard, it is  not unexpected to observe a resemblance with the fixed-order QCD + PS calculation. Second, there is a competition between the growth of the BFKL Green's function with larger rapidity differences between the MN jets and the decrease of the parton distribution functions with increasing $x$. This effectively introduces a sector of the possible energy-momentum conserving terms (a global one), which is not related to higher orders in the BFKL formalism. To illustrate this point further, we have a more detailed look at the normalized differential in $\Y$ cross section in Fig.~\ref{dSigmadYfull}, where we observe that for the larger values of $\Y$ the BFKL cross section is more suppressed than the fixed-order QCD one. The differences between both calculations are more pronounced at smaller  $p_{T,{\rm min}}$, as in the displayed plot, since this is the case where the BFKL dynamics is less affected by this cutoff.

\subsection{Effect of multiple parton interactions}

\begin{figure}
\begin{center}
\includegraphics[width=9.cm]{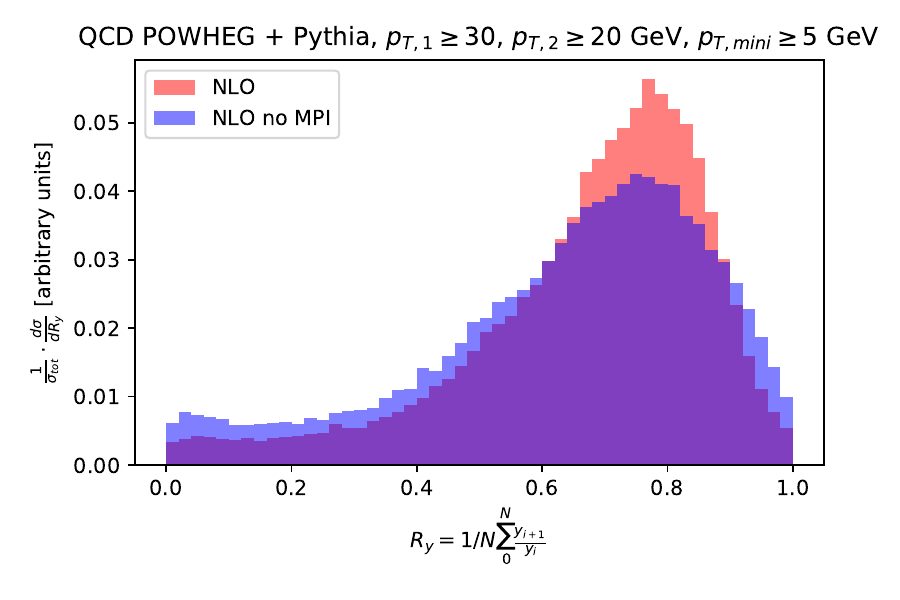}\\
\vspace{-.3cm}
\includegraphics[width=9.cm]{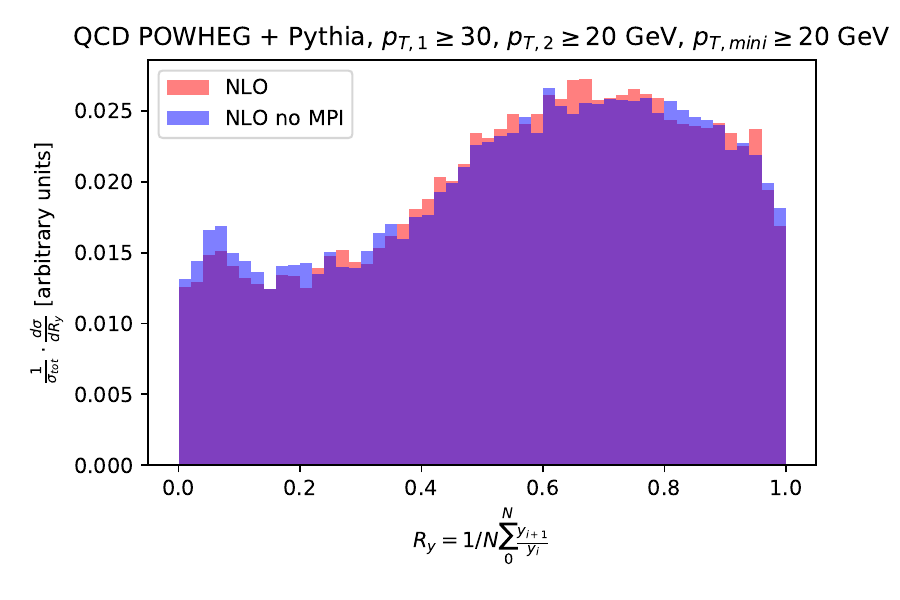}
\caption{Comparison of $R_y$ with and without MPI for minijet with cutoffs $\pM=\SI{5}{\giga\electronvolt}$ (top) and $\pM=\SI{20}{\giga\electronvolt}$ (bottom). The effect of MPI is largely reduced with the higher $\pM$ requirement.}
\label{MPI}
\end{center}
\end{figure}

The observable relies on the use of events with high jet multiplicity, where the jets have relatively low $\pT$ and span the whole detector acceptance in rapidity. This makes the event shape observables potentially sensitive to MPI contributions, which are not related to the emergent multi-Regge kinematics of interest for the hardest parton-parton scattering. We use the $R_y$ observable to illustrate the effect of MPI on the proposed observables (similar conclusions are found for the other observables). The presence of MPI increases the jet multiplicity, which can be seen in  the top panel of Fig.~\ref{MPI}. To avoid double counting, the MPI in \textsc{PYTHIA}8 starts at a maximum $\pT$ of the factorization scale, so it produces mostly softer jets than the hard process. This is seen in the Fig.~\ref{MPI} lower, where no differences are found when increasing $\pM$ from $\SI{5}{\giga\electronvolt}$ to $\SI{20}{\giga\electronvolt}$. The presence of MPI should be taken into account when considering minijets should be considered for analyses with a lower jet $\pT$ threshold. This conclusion applies to the other event shape observables studied here, and thus justifies neglecting MPI for the present (for the lower $\pT$ thresholds accessible by ATLAS and CMS).

\section{Summary}

In this paper, we have focused in multijet production in the high-energy limit of quantum chromodynamics (QCD) using events with the Mueller--Navelet (MN) multijet topology. Such a topology is expected to be sensitive to the emergent properties of the strong interaction expected from the high-energy limit of QCD, described by the Balitsky--Fadin--Kuraev--Lipatov (BFKL) evolution equations of QCD. We have investigated a set of event shape observables at fixed jet multiplicities that could be measured at the LHC. The observables are used to compare two set of predictions: one of them is based on fixed-order next-to-leading order perturbative QCD calculations supplemented with a parton shower based on Dokshitzer--Gribov--Lipatov--Altarelli--Parisi (DGLAP) evolution at leading-logarithmic accuracy simulated with POWHEG+PYTHIA8 (dubbed ``NLO+PS''), representative of the standard Monte Carlo simulation of multijet events at the LHC, and the second set of predictions is based on the BFKL formalism at leading-logarithmic accuracy using the {\tt BFKLex} Monte Carlo event generator, which accounts for the resummation of multiple gluon emissions in the high-energy limit of QCD (dubbed ``BFKL LL'').\\

Three observables were introduced, which have different sensitivities to the multijet topologies of interest: the rapidity ratio $R_y$, the mean rapidity difference $\yobs$ between minijets  and the invariant mass ratio $R_{ky}$, whereas we also studied the average minijet transverse momentum $\langle \pT \rangle_{\text{mini}}$.\\

For realistic event selection requirements for CMS and ATLAS detectors' acceptance and the fixed jet multiplicities that we considered, we observe minor differences in the predictions from both NLO+PS and BFKL LL approaches. This is encouraging from a theoretical perspective, since it demonstrates the reliability of the BFKL formalism to describe multijet production for processes with a relatively low center-of-mass energy relative relative to other momentum scales and for more less inclusive jet observables. For low minijet multiplicity events, the BFKL predictions produce very similar results to the NLO+PS approach. However, we should emphasize that, due to experimental limitations, it is always necessary to exclude low transverse momenta contributions from the BFKL gluon Green's function in order to compare to very exclusive experimental data. In this way, we are always testing the hardest tail of the minijet contributions to the Green's function. It is well-known that those untagged regions of phase space are crucial in the BFKL formalism, since it corresponds to the gluon emissions that yield the largest corrections.\\

The difference in the predictions for $\pobs$ can be attributed to features of the employed approximations, and cannot be deemed relatable to any experimental data. The conclusion is that, with the event shape variables presented in this work, it is challenging to cleanly isolate BFKL-like dynamics from DGLAP-like dynamics in the kinematical regime of the LHC. These findings motivate a measurement of the observables to constrain the modeling of multijet production, to either confirm the findings obtained in the simulations, or to inspire further development towards yet more accurate Monte Carlo methods for this kinematic regime.\\

On the experimental side, these topologies require very low $\pT$ jets at central rapidities. The setup used here is based on the jet $\pT$ limitations of the ATLAS and CMS Collaborations using calorimetric and track information in their jet finding algorithm and in the jet energy calibration, which allows to use calibrated jets at $\pT \approx$ 20 GeV. An alternative setup of these observables could be to consider charged-particle jets for the minijet activity, which could allow to further extend the low $\pT$ reach significantly. The extent to which such low $\pT$ jets bear a faithful correspondence to the particle-level minijets is something that could be investigated by the experiments. Also, considering that the MN jet topology is very specific (one very forward jet, one very backward jet), one can think of designing dedicated dijet triggers that can operate during low pileup runs, for example possible runs planned for the $W$ boson mass measurement at the LHC. Another possibility for reaching lower $\pT$ is to use identified hadrons (e.g., heavy-flavor hadrons) in association with a forward jet.\\

The current study opens up opportunities for future work. The {\tt BFKLex} generator clearly allows for the study of more exclusive observables within the BFKL formalism, where the analytical approach becomes less straightforward if not impossible. Our next objective involves employing a complete NLL BFKL gluon Green's function convoluted with NLO jet vertices, a task that presents significant challenges. Additionally, exploring the impact of incorporating parton showering effects on the distributions obtained from a BFKL-based approach holds considerable interest. We anticipate sharing our progress in a forthcoming publication.

\section*{Acknowledgements}

This work has been supported by the German Research Foundation (DFG) through the Research Training Network 2149 ``Strong and weak interactions - from hadrons to dark matter'', by the Spanish Research Agency (Agencia Estatal de Investigaci{\'o}n) through the grant IFT Centro de Excelencia Severo Ochoa SEV-2016-0597, and by the Spanish Government grant FPA2016-78022-P. It has also received funding from the European Union’s Horizon 2020 research and innovation programme under grant agreement No.\ 824093. The work of GC was supported by the Funda\c{c}{\~ a}o para a Ci{\^ e}ncia e a Tecnologia (Portugal) under project EXPL/FIS-PAR/1195/2021 (http://doi.org/10.54499/EXPL/FIS-PAR/1195/2021) and contract ‘Investigador FCT - Individual Call/03216/2017’. JGM was partly supported by the European Research Council project ERC-2018- ADG-835105 YoctoLHC. GC and JGM were also supported by Funda\c{c}{\~ a}o para a Ci{\^ e}ncia e a Tecnologia (Portugal) under project CERN/FIS-PAR/0032/2021 (http://doi.org/10.54499/CERN/FIS-PAR/0032/2021). CB is supported by the European Research Council consolidator grant no. 101002207 (\textit{QCDHighDensityCMS}).

\bibliographystyle{ieeetr}

\bibliography{refs}

\end{document}